\documentclass[reprint,amsmath,amssymb,aps,prl,floatfix,superscriptaddress,twocolumn]{revtex4-2}

\newcommand{\env}[1]{\texttt{#1}}
\usepackage{graphicx}
\usepackage{docs}%
\usepackage{amsmath}
\usepackage{amsfonts}
\usepackage{comment}
\usepackage[colorlinks=true,linkcolor=blue]{hyperref}
\usepackage[capitalize]{cleveref}

\usepackage{float}

\makeatletter
\def\maketitle{
\@author@finish
\title@column\titleblock@produce
\suppressfloats[t]}
\makeatother

\begin{document}

\title{Flow fields around active droplets squeezing through tight confinements}
\author{Subhasish Guchhait}%
\email{The authors have contributed equally to this work}
\affiliation{
 Department of Mechanical and Aerospace Engineering,\\
Indian Institute of Technology Hyderabad,
Kandi, Sangareddy 502285, India }
\author{Smita Sontakke}%
\email{The authors have contributed equally to this work}
\affiliation{
 Department of Mechanical and Aerospace Engineering,\\
Indian Institute of Technology Hyderabad,
Kandi, Sangareddy 502285, India }
\author{Shubhadeep Mandal}%
\affiliation{
Department of Mechanical Engineering,\\
Indian Institute of Science,
Bengaluru 560012, India }
\author{Ranabir Dey}%
\email{ranabir@mae.iith.ac.in}
\affiliation{ 
Department of Mechanical and Aerospace Engineering,\\
Indian Institute of Technology Hyderabad,
Kandi, Sangareddy 502285, India }
%

\begin{abstract}
Biological microswimmers, like euglena, deform their body shape to swim through tight confinements having length scales comparable to the microswimmer length scale.
Recently, it was shown that self-propelling active droplets can also squeeze through tight microconfinements by elongating their shape.
However, the evolution of the hydrodynamic signature, or the velocity field, generated by the active droplet, as it deforms its shape to swim through increasingly tight microconfinements, has remained scarcely studied.
Using high-resolution fluorescence microscopy and $\mu$-Particle Image Velocimetry (PIV) analysis, we show here that as the swimming active droplet deforms from a spherical shape to a `stadium'-like shape, and eventually to an elongated `capsule'-like shape in increasingly tighter microchannels, its hydrodynamic signature changes from a `pusher'-like velocity field to a `puller'-like velocity field, and finally to an asymmetric velocity field. 
We characterize such alterations in the active droplet dynamics using the distributions of the local velocity magnitude, axial and transverse components of the local flow velocity, vorticity, and the filled micelle concentration. 
Finally, we use finite-element based numerical simulations to explain the aforementioned evolution of the velocity field stemming from the underlying physico-chemical hydrodynamics in the presence of a thin lubrication film between the swimming active droplet and the tight microconfinement walls.
The present work provides a comprehensive understanding of the  chemo-hydrodynamic characteristics of active droplets navigating extreme confined spaces, which can be beneficial for many autonomous cargo-delivery applications. 
\end{abstract}
\maketitle

\section{Introduction}
Motile microorganisms like algae, bacteria, and protozoa, e.g. \textit{paramecium}, are typically found in diverse environment such as in natural water bodies \cite{Microorganism_lake, Microorganism_Sea}, in soil \cite{Bacteria_in_soil_Ranjard, Bacteria_in_soil_Smets}, and even within the human body\cite{Microorganism_humanbody,Microorganism_humanbody_2}.
For their survival in complex environment, these biological microswimmers need to navigate through confined spaces, like in soil \cite{Bacteria_in_soil_Ranjard, Bacteria_in_soil_Smets, Bacteria_in_soil_Jennifer} and through mucus network \cite{Bateria_in_mucus, COHEN1995309, Bateria_penetrate_mucus, Bateria_penetrate_colon_mucus}, and adopt various swimming strategies for efficient uptake of oxygen and nutrients. 
To understand these survival strategies of biological microswimmers, it is then important to understand how confined spaces alter the swimming dynamics of microswimmers.

Some studies on \textit{E. coli} in microchannels indicate that these bacteria tend to swim near corners and walls in wider channels \cite{Ecoli1Dswimming, BacterialmotilityPNAS}. 
Their swimming velocity increases, reaching a peak value as the channel cross-section decreases, and they eventually switch to swimming along the axis of the channel \cite{Ecoli1Dswimming}. 
In even tighter confinements, swimming velocity of \textit{E. coli} decreases, and they navigate through the channel by growing and dividing their cells \cite{Bacterialgrowth_PNAS}. 
\textit{Euglena} \cite{Euglena} and \textit{paramecium} \cite{Paramecium, Paramecium2} can also swim through a tight confinement by deforming their body shape and by adopting a cylindrical shape respectively. 
During such swimming in tight confinements, these microswimmers never come in direct contact with the walls of the confinement, which may be attributed to the existence of a thin lubricating film between the wall and the body of the microswimmer. 
The role of an intervening thin film in altering the swimming characteristics of microswimmers in tight confinements remains scarcely studied.
Such studies can subsequently aid in designing applications in which self-propelling, artificial microswimmers need to navigate tight confinements, e.g. in targeted cargo delivery for biomedical applications \cite{medina2018micro,lin2021self}.

Over the last decade or so, researchers have designed self-propelling artificial microswimmers which mimic the motility, and often the hydrodynamic signature, of biological microswimmers.
These self-propelling artificial microswimmers can be of two types -- first, there are the Janus particles \cite{paxton2004catalytic, howse2007self, zottl2016_emergent, liebchen2021_interactions, zottl2023_modeling} having an engineered material asymmetry.
Janus colloidal particles self-propel due to phoretic effects \cite{anderson1989colloid, moran2017phoretic} stemming from catalytic reactions, such as self-diffusiophoresis \cite{self-diffusiophoresis}, self-electrophoresis \cite{pumera2010electrochemically}, or self-thermophoresis \cite{jiang2010active, yu2019phototaxis}. 
Second, there are the isotropic active oil/water droplets \cite{toyota2009_selfpropelled, Swarmingbehavior2011, peddireddy2012solubilization, herminghaus2014interfacial, izri2014self, suga2018self, Swimmingdroplets2016, birrer2022_we, dwivedi2022_selfpropelled, michelin2023_selfpropulsion} which self-propel in surfactant solutions by generating a surface tension gradient along the interface due to spontaneous symmetry breaking of the interfacial surfactant monomer concentration.
From the perspective of applications, active droplets have certain advantages over active catalytic colloids-- (i) the former being essentially isotropic microscale oil/water droplets can be generated on-demand using standard microfluidic techniques by circumventing involved fabrication methodologies necessary for engineering Janus colloids; (ii) active droplets use ionic surfactant solutions as their fuel which can be chosen for their biocompatibility unlike the corrosive chemicals, like hydrogen peroxide, necessary as fuels for active colloids.
Hence, it is logical to conclude that active droplets can be a more viable alternative as autonomous micro-carriers for applications like targeted cargo delivery in biomedical settings.
However, in applications like targeted cargo delivery, both active droplets and catalytic Janus colloids, more often than not, need to navigate narrow confinements having length scales typically of the order of the swimmer length scale \cite{xiao2018review}. 

Over the years, substantial theoretical efforts have been made to understand the effects of microchannel/microcapillary geometry, cross-section, and size on the dynamics of self-propelled microswimmers in confinements \cite{Capillary, Elgeti, Graaf, geometry, Ahana_FDR, dhar2020hydrodynamics}.
On the experimental front, it was demonstrated that catalytic Janus colloids are attracted towards, and tend to swim along, solid boundaries \cite{simmchen2016topographical} -- a proclivity which can be exploited to guide these self-propelled microswimmers towards a desired target.
Furthermore, catalytic Janus microswimmers also demonstrated a substantial increase in their swimming speed with increasing tightness of the confinement \cite{liu2016bimetallic}.
In case of active droplets, investigations were conducted to delineate their swimming characteristics near a confining wall \cite{de2019flow}, in complex microfluidic geometries \cite{jin2017chemotaxis, jin2018chemotactic}, and even in the midst of micro-pillar arrays \cite{jin2018chemotactic,jin2019fine}.
Fascinatingly, active droplets can `sense' surfactant concentration gradients enabling them to swim through complex microfluidic networks in a manner reminiscent of auto-chemotactic swimming strategies \cite{jin2017chemotaxis, jin2018chemotactic}.
Near a micro-pillar, active droplets exhibit an attraction towards, or repulsion away from, a pillar depending on the radius of curvature \cite{jin2019fine}. 
Recently, efforts were also made to delineate the emergent interfacial hydrodynamics of immobilized active droplets behaving as micropumps \cite{ramesh2023interfacial}, and to understand how the underlying physico-chemical hydrodynamics shapes mutual interactions among active droplets in confinements \cite{kumar2024motility}.
At this point, it is interesting to wonder whether active droplets can autonomously squeeze through tight microchannels having length scales smaller than the microswimmer length scale by deforming their shape, in a manner analogous to biological microswimmers, like \textit{euglena} \cite{Euglena}, and \textit{paramecium} \cite{Paramecium, Paramecium2}. 

In an interesting recent work \cite{de2021swimming}, it was demonstrated that active water droplets can indeed squeeze through tight microchannels, straight or converging-diverging, by elongating their shape. 
With increasing confinement, the elongated active droplets undergo spontaneous division at their posterior end \cite{de2021swimming}.
The aforementioned behaviour of active droplets was explained by considering the chemo-hydrodynamics in presence of a surrounding lubricating thin film, in a manner similar to the classical Bretherton approach \cite{de2021swimming}.
In a subsequent theoretical work \cite{francesco2022confined}, the self-propulsion of an isotropic, autophoretic spherical particle in a tight capillary was also explained, with a special focus on the physico-chemical hydrodynamics considering the thin lubrication layer.
Except for a brief qualitative effort in \cite{de2021swimming}, the evolution of the hydrodynamic signature (flow field) of an active droplet as it spontaneously adapts its shape to swim through increasingly tighter confinement has remained unstudied.
Such an investigation will provide a clear understanding of how the coupling of the shape (geometry) of the microswimmer and the underlying physico-chemical hydrodynamics, in presence of a thin lubricating film, alters the velocity field as the active droplet autonomously squeezes through a tight confinement. 

In this work, we explain the evolution in the velocity field generated by increasingly bigger self-propelled oil droplets as they squeeze through  a tight rectangular microchannel using high-resolution fluorescence microscopy and quantitative micro-Particle Image Velocimetry ($\mu$-PIV) analysis.
We show that as the active droplet becomes increasingly bigger relative to the tight microchannel, it adapts its shape from spherical to `stadium'-like, and finally to an elongated `capsule'-like shape in order to spontaneously squeeze through the microconfinement.
Interestingly, during this shape evolution, the hydrodynamic signature of the active droplet changes from a `pusher'-like flow field, to a `puller'-like flow field, and eventually to an asymmetric velocity field characterized by two circulation zones at the anterior and posterior. 
To better understand the dynamics of the self-propelling active droplet, we also perform fluorescence microscopy analysis to delineate the ejected filled micelle concentration around the active droplet corresponding to the varying shapes in the tight microchannel.
Finally, we complement our experimental observations with a full-scale numerical simulation of the coupled hydrodynamic and the surfactant advection-diffusion problem for the active droplet of different shapes in a tight microchannel. 
Using the understanding developed from the simulation results, we try to explain how the shape of the active droplet in the tight microconfinement and the surfactant transport characteristics through the intervening thin lubrication film, between the microswimmer and the confinement walls, dictate the resulting hydrodynamic signature.
We think that the present work is going to aid in a comprehensive understanding of the dynamics, and of the ensuing physico-chemical hydrodynamics, of self-propelling active droplets squeezing through tight confinements.

To explain the aforementioned techniques and results, this article is structured as follows: in section II, we describe in detail the various experimental setups and methodologies; in section III, we describe the details of the numerical simulation technique; in section IV, we present and discuss the results obtained through the various experimental techniques; in section V, we explain the experimental observations with the help of the numerical simulation results; and in section VI, we summarize our findings, and discuss their consequences for future endeavours. 

\section{Experimental setup and methodology}
\subsection*{Fabrication of PDMS microchannels}
We use the well-established photolithography and soft-lithography techniques to fabricate the microchannels used in this work. We pour polydimethylsiloxane (PDMS) mixture, consisting of Sylgard 184 prepolymer and cross-linker in the weight ratio of $10:1$, onto the silicon wafers containing microfluidic channel moulds fabricated using SU8 photoresist. 
Subsequently, we cure the PDMS at $75^\circ$C in a hot air oven for a duration of 3 hours.
Using this procedure, we fabricate two types of microfluidic chips: (i) a flow-focusing cross junction for generating active droplets, and (ii) microfluidic chips containing multiple channels connected to single inlet and outlet reservoirs (see Supplementary Material (SM)-Fig.~\ref{fgr:10}).  
Each microchannel has a height of $h =$ $50$ $\mu m$ and a width of $w =$ $64$ $\mu m$ (Fig.~\ref{fgr:1}(a)). 
Finally, the cured PDMS stamps are bonded to PDMS-coated coverslips (thickness: 160 $\mu m$) using plasma oxidation. 
\subsection*{Generation of active droplets}
We use slowly solubilizing CB15 oil ((S)-4-Cyano-4$'$-(2-methyl butyl) biphenyl; Tokiyo Chemical Industry) droplets in a supramicellar aqueous solution of cationic surfactant Tetradecyltrimethylammonium bromide, TTAB (Critical micelle concentration (CMC) = 0.13 wt.$\%$; Sigma-Aldrich) as the active droplet system \cite{peddireddy2012solubilization,herminghaus2014interfacial, Swimmingdroplets2016,michelin2023_selfpropulsion}. 
First, we generate CB15 droplets in 0.1 wt$\%$ aqueous surfactant solution using the flow-focusing cross-junction chip (see SM-Fig.~\ref{fgr:10}).
We use four different sizes (diameter $D_d$) of the droplets for the experiments reported here -- $45.33$ $\pm$ $3.01$ $\mu m$; $55.72$ $\pm$ $3.89$ $\mu m$; $67.6$ $\pm$ $2.86$ $\mu m$; and $80.64$ $\pm$ $6.39$ $\mu m$.
We make these droplets active, i.e. self-propelling, by mixing the $0.1$ wt$\%$ aqueous surfactant solution (containing monodispersed droplets) in 7.5 wt$\%$ aqueous surfactant solution (surfactant concentration above CMC) in a  $1:2$ ratio. 

These active droplets self-propel due to micellar solubilization \cite{peddireddy2012solubilization,herminghaus2014interfacial,Swimmingdroplets2016,michelin2023_selfpropulsion}. 
During this solubilization process, empty micelles in the aqueous solution take up oil molecules and additional surfactant monomers from the interfacial region, and grow into filled micelles. 
Due to any advective perturbation, the initially isotropic distribution of filled and empty micelles around the droplet gets spontaneously broken, resulting in local enhancement (reduction) in filled to empty micelle ratio along the interfacial region. 
This locally reduces (enhances) the interfacial surfactant monomer coverage, setting up an interfacial tension variation inversely proportional to the local surfactant monomer coverage. 
Such spontaneously created interfacial tension gradient along the droplet interface in turn sets up a local Marangoni stress.
This Marangoni stress triggers an interfacial velocity towards the region of lower surfactant coverage (away from the region of higher empty micelle concentration and towards the region of higher filled micelle concentration).
To satisfy the non-inertial constraints of the system, the oil droplet starts to self-propel in the direction of higher concentration of empty micelles, thereby maintaining a self-sustaining interfacial tension gradient.
As the active droplets swim in a quasi-ballistic manner, they leave behind a trail of filled micelles in their wake \cite{hokmabad2021emergence, hokmabad2022chemotactic}. 

\subsection*{Bright field microscopy}
\begin{figure}[h!]
 \centering
 \includegraphics[width=8.2cm]{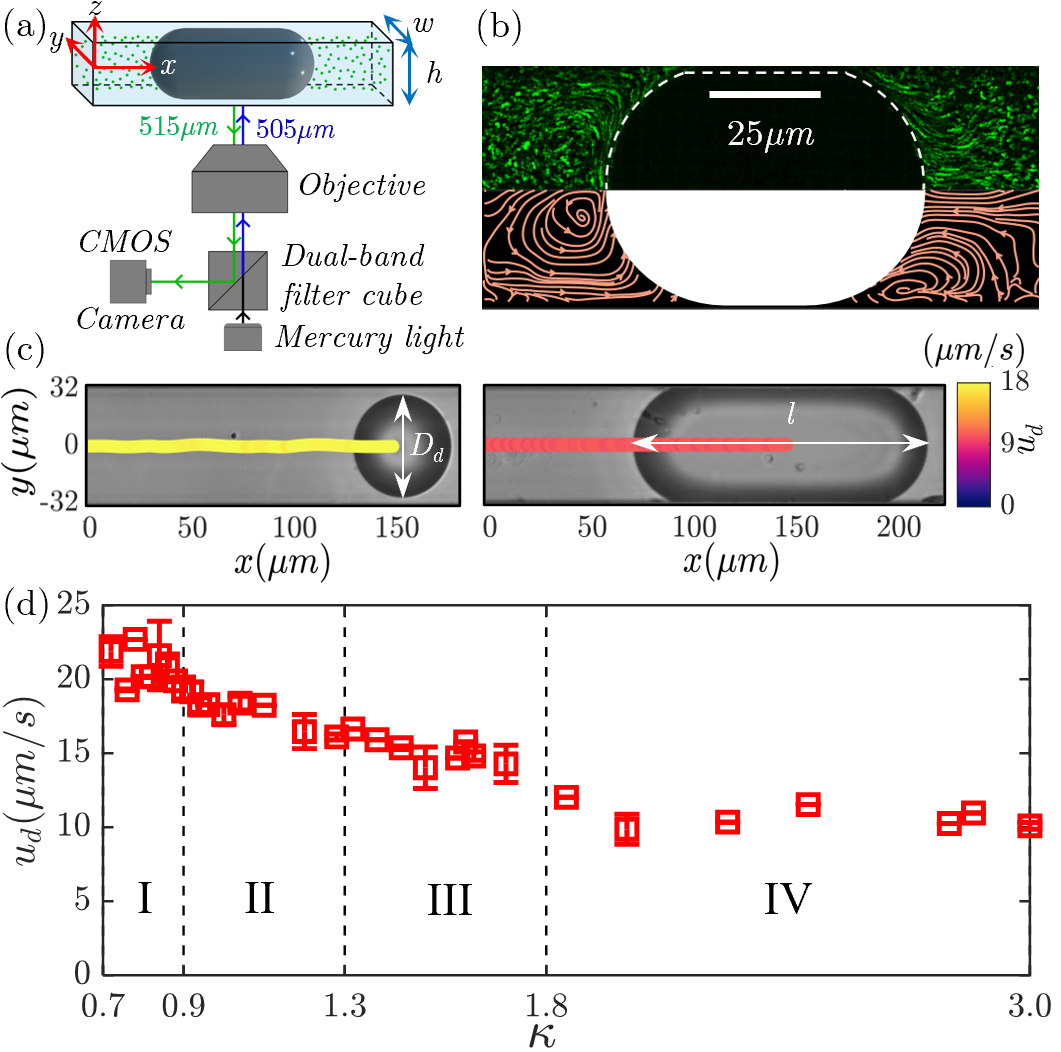}
 \caption{\label{fgr:1}\textbf{(a)} Schematic of an elongated active droplet inside a rectangular microchannel, with the microscopy setup. \textbf{(b)} A fluorescence micrograph depicting the flow around a self-propelling active droplet (for $\kappa=\frac{D_d}{D_h}$ or $\frac{l}{D_h}$ $=1.50$) (top half), and the corresponding streamline pattern obtained using $\mu$PIV analysis (bottom half).\textbf{(c)} Bright field microscopy images showing the active droplet trajectory inside the microchannel color-coded with the swimming velocity $u_d$ of the droplet for (left) spherical ($\kappa<$ $1.00$) and (right) elongated ($\kappa=3$) droplets.\textbf{(d)} Variation of the steady droplet swimming velocity with increasing confinement effect ($\kappa$).}
\end{figure}

We use bright field microscopy (with an inverted Nikon ECLIPSE Ti microscope) at $30\times$ magnification to characterize the trajectories of the self-propelled active droplets of varying sizes in the microchannel with a fixed rectangular cross-section. 
The extent to which the active droplet is confined in the microchannel is characterized by the non-dimensional parameter $\kappa$, where for spherical droplets $\kappa=D_{d}/D_{h}$, and for elongated droplets $\kappa=l/D_{h}$. Here, $D_d$ is the diameter of spherical droplet, $l$ is the maximum length of the elongated active droplet, and $D_{h}$ is the hydraulic diameter of the channel defined as $4wh/(2(w+h))=$ $56.14$ $\mu m$.
Higher values of $\kappa$ indicate a tighter confinement relative to the characteristic length scale of the droplet. 
For $\kappa < 1.30$, the active droplet swims in the microchannel with a spherical or nearly spherical shape (Fig.~\ref{fgr:1}(c) left). 
For $1.30 < \kappa \leq 1.80$, the swimming droplets adopt a stadium-like shape \cite{Stadium} inside the microchannel (Fig.~\ref{fgr:1}(b)). 
Finally, the droplets exhibit an elongated shape, like a capsule \cite{Capsule}, for $\kappa>$ $1.80$ as they swim in the microchannel (Fig.~\ref{fgr:1}(c) right).
For $\kappa<$ $1.30$, the active droplets spontaneously enter the microchannel from the reservoir in which they are dispersed. 
However, for $\kappa>$ $1.30$, we `inject' the bigger active droplets into the microchannel using an imposed ambient flow. Subsequently, we wait till the ambient reaches a quiescent state before recording any image.

We record the videos at a frame capture rate of 25 fps (frame rate per second) using a CMOS monochrome digital camera (IDS Imaging) having $1920 \times 1200$ pixels in the $1/1.2^{\prime\prime}$ CMOS sensor.
We determine the swimming trajectory, and calculate the instantaneous swimming velocity $(u_d)$ of the active droplet by tracking the centroid of the droplet projection in the microscopy image plane using an in-house image processing routine (in MATLAB R2020a) (Fig.~\ref{fgr:1}(c)) (also see SM, section-II). 
For the experiments reported herein, the droplets exhibit steady-state motion (Fig.~\ref{fgr:1}(c)).
Hence, the measured instantaneous velocity of the droplets represents their steady swimming velocity. 
The minimum translation velocity that we can resolve using a $30\times$ objective (with a spatial resolution of $0.176$ $\mu m/pixel$) and a frame capture rate of 25 fps is $2.2$ $\mu m /s$. 

\subsection*{Fluorescence microscopy and micro-Particle Image Velocimetry ($\mu$PIV)}
For varying $\kappa$, the velocity field generated by the active droplet, during swimming in the microchannel, is characterized using high-resolution fluorescence microscopy, and subsequent micro-Particle Image Velocimetry ($\mu$PIV) analysis. 
The $\mu$PIV setup, consisting of the illumination modules and the dual-band filter cube (peak excitation and emission wavelengths: 487/562 $nm$ and 523/630 $nm$), is integrated with the inverted Nikon ECLIPSE Ti microscope (Fig.~\ref{fgr:1}(a)). 
To visualize the flow field, we use 500 $nm$ carboxylate-modified fluorescent microspheres (Thermo Fisher Scientific) having excitation/emission wavelengths of $505/515$ $nm$. 
We carry out the subsequent analysis using the open-source PIVlab \cite{stamhuis2014pivlab}, and our in-house post-processing routine. 

For the $\mu$PIV analysis using PIVlab, we use the Discrete Fourier transform (DFT) method to get the cross-correlation matrix. 
We use multiple passes with $64 \times 64$ pixels interrogation area with a step size of $32 \times 32$ pixels, followed by a second pass of $32 \times 32$ pixels interrogation area with a step size of $16 \times 16$ pixels.
We then extract the data sets containing the velocity vector field from PIVlab, and post-process these using an in-house MATLAB routine to generate the velocity contour plots and streamlines (Figs.~\ref{fgr:2} and \ref{fgr:3}), and the vorticity contour plots (Fig.~\ref{fgr:4}) (see Appendix, section A1 (Fig.~\ref{fgr:8}) for other details). The average advective velocity around the droplet for $\kappa = 0.85$ (Fig.~\ref{fgr:2}(a)) at a distance equal to the radius of the droplet ($\sim D_{d}/2$)  from its circumference estimated from the $\mu$PIV analysis is $7.169 \pm 0.3312$ $\mu m/s$.
In Fig.~\ref{fgr:1}(b), a fluorescence micrograph depicting the flow field around an active droplet is shown in the top half, and the corresponding streamlines obtained from $\mu$PIV analysis are shown in the bottom half.

\subsection*{Filled micelle trail visualization and kymographs}
To visualize the filled micelle trail associated with the droplet microswimmer during its self-propulsion \cite{hokmabad2021emergence, hokmabad2022chemotactic}, we perform a separate set of experiments under identical conditions.
In this case, we dye the CB15 oil with fluorescent Nile Red dye (Invitrogen™) (excitation/emmission: $550/640$ $nm$), and generate the oil droplets using similar droplet generation methodology. 
We then mix these Nile Red doped droplets in $7.5 \%$ aqueous TTAB solution to make them active, and introduce them in the microchannels. 
We use the identical inverted microscope setup with the dual band filter to visualize the trail of filled micelles containing the doped oil molecules which now fluoresce. 
During the image recording (at 25 fps), we adjust the exposure such that the filled micelle trail is clearly visible without over-saturating the pixels around the swimming droplet. 
We extract the fluorescence signal intensity ($I$) along the droplet interface precisely at a distance of $14$ $\mu m$ outside its circumference using a separate in-house Matlab code. 
 
At each instant of time over a propulsion period of about 20 s, we extract the pixel intensity $I$ along the droplet interface (always at a distance of about $14$ $\mu m$ outside the interface) by scanning following a clockwise rotation about the droplet centroid, starting from the x-axis which is aligned with the channel center-line (Fig.~\ref{fgr:5}).
The angular $(\theta)$ position [0, 2$\pi$] represents the anterior apex of the active droplet aligned with its swimming direction, while the angular position ($\theta=\pi$) represents the diametrically opposite posterior apex of the swimming droplet.
We plot the temporal variation of $I$ along the circumference from [0, 2$\pi$] over a period of about 20 s to generate the kymographs shown in Fig.~\ref{fgr:5} (right column). 
Here, the intensity values are non-dimensionalized by the maximum intensity value ($I_{max}$) among all the experiments.
We repeat identical experiments and post-processing methodology to generate the kymographs for droplets swimming with increasing values of $\kappa$.

\section{Numerical simulation}
\subsection*{Governing equations and boundary conditions}
The physico-chemical hydrodynamics of chemically active droplets is governed by the nonlinear coupling between fluid flow and chemical transport of surfactant monomer, filled/swollen micelles and empty micelles. We simplify the physical system by using a reference model \cite{michelin2013spontaneous, morozov2019nonlinear, michelin2023_selfpropulsion} in which the chemical transport outside the droplet phase is modelled as a simple advection-diffusion process of surfactant molecules that is consumed at the droplet interface at a fixed rate. The velocity $(\boldsymbol{u})$ and pressure $(p)$ fields are governed by the continuity and Navier-Stokes equations of the form \cite{michelin2023_selfpropulsion}

\begin{subequations}
    \begin{align}
        \nabla \cdot \boldsymbol{u}_{i,e} &= 0,  \\
        \rho_{i,e} \left( \frac{\partial \boldsymbol{u}_{i,e}}{\partial t} + \boldsymbol{u}_{i,e}\cdot \boldsymbol{\nabla} \boldsymbol{u}_{i,e} \right) &= -\boldsymbol{\nabla}p_{i,e} + \mu_{i,e} \nabla \boldsymbol{u}_{i,e}, 
        \label{FlowEq}
    \end{align}
\end{subequations}

\noindent where $\rho$ is the density and $\mu$ is the dynamic viscosity. Subscripts $i$ and $e$ denote droplet phase and suspending phase, respectively. At the fluid-fluid interface, the interacting fluids satisfy velocity continuity and kinematic conditions 
\begin{subequations}
    \begin{align}
        \boldsymbol{u}_{i} &= \boldsymbol{u}_{e},  \\
        \frac{d \boldsymbol{x}_s}{dt} &= \left(\boldsymbol{u}_{e} \cdot \boldsymbol{n} \right) \boldsymbol{n},
        \label{VelBC}
    \end{align}
\end{subequations}
\noindent where $\boldsymbol{x}_s$ represent the interface position and $\boldsymbol{n}$ represents the outward unit normal (directed away from the droplet phase). The stress balance at the fluid-fluid interface is given by \cite{morozov2019self}
\begin{equation}
    \boldsymbol{n} \cdot (\boldsymbol{\sigma}_{i} - \boldsymbol{\sigma}_{e}) = \gamma (\boldsymbol{\nabla}_s \cdot \boldsymbol{n})\boldsymbol{n} - \boldsymbol{\nabla}_s \gamma, 
    \label{StressBC}
\end{equation}

\noindent where $\boldsymbol{\sigma}$ is stress tensor and $\gamma$ is the surface tension. It is important to note that for chemically active droplets the fluid flow is coupled to the surfactant transport via the interfacial Marangoni stress (represented by the gradient of surface tension term in stress balance equation). The surface tension depends on the surfactant concentration as \cite{morozov2019self} $\gamma = \gamma_0 - \gamma_c (C - C_{\infty} + AR/2D)$, where $C$ is the surfactant concentration, $C_{\infty}$ is the far-field surfactant concentration and $R$ is the length scale. Here, $\gamma_0$ and $\gamma_c$ are positive constants signifying reference surface tension at $C = C_{\infty} - AR/2D$ and $\gamma_c \equiv (d\gamma/dC)|_{C=C_{\infty}-AR/2D}$, respectively. The surfactant concentration outside the droplet is governed by an advection-diffusion equation of the form \cite{michelin2023_selfpropulsion}
\begin{equation}
    \frac{\partial C}{\partial t} + \boldsymbol{u}_e \cdot \boldsymbol{\nabla}C = D \nabla^2 C,
    \label{CEq}
\end{equation}

\noindent where $D$ is the molecular diffusivity of surfactant. As the droplet consumes surfactants at a constant rate, the diffusion flux at the droplet interface is given as: 
\begin{equation}
    D \boldsymbol{n} \cdot \boldsymbol{\nabla} C = A,
    \label{CBC1}
\end{equation}

\noindent where $A$ $(>0)$ is the constant rate of consumption. The bounding walls do not consume surfactants, thus the boundary condition at the walls are no-slip and no-penetration for velocity field, and no-flux condition for surfactant concentration:
\begin{equation}
    D \boldsymbol{n}_w \cdot \boldsymbol{\nabla} C = 0,
    \label{CBC2}
\end{equation}
\noindent where $\boldsymbol{n}_w$ represents the outward unit normal (directed away from the wall).

We have non-dimensionalized the mathematical model using the following characteristic scales: confinement length scale $R$ as the characteristic length scale, $AR/D$ as the concentration scale, $\mathcal{U} = A\gamma_c R/[D(2\mu_e + 3\mu_i)]$ as the velocity scale, $R/\mathcal{U}$ as the time scale, and $\mu_e \mathcal{U}/R$ as the pressure scale. This non-dimensionalization scheme yields the following dimensionless numbers: Reynolds number $Re=\rho_e \mathcal{U} R/\mu_e$, capillary number $Ca=\mu_e \mathcal{U}/\gamma_o$, P\'eclet number $Pe= \mathcal{U} R/D$, confinement ratio $\kappa = D_d/2R$, density ratio $\rho_i/\rho_e$ and viscosity ratio $\mu_i/\mu_e$. The non-dimensional variables obtained from the numerical simulation are represented with asterisk as superscript.

\subsection*{Solution methodology}
To perform the numerical simulations, we have solved the governing differential equations for surfactant concentration, velocity and pressure fields along with boundary conditions (Eq.(1) -- Eq.(6)) using the finite-element solver COMSOL Multiphysics. We have used the ‘Laminar Flow’, Convection-Diffusion Equation’ and  ‘Moving Mesh’ modules of COMSOL Multiphysics to model the coupled nonlinear system. The ‘Moving Mesh’ module employs the arbitrary Lagrangian-Eulerian (ALE) method \cite{hadikhani2018inertial, yu2018time, balestra2018viscous, mao2023effect} along with automatic remeshing algorithm to capture the transient dynamics of the moving and deforming fluid-fluid interface. All the simulations are performed in a two-dimensional cylindrical domain of radius $R$ and length $30R$. To avoid large deformation of fluid-fluid interface, we solve the problem in a co-moving reference frame in which reference frame translates with the droplet along the channel axis. In this co-moving frame, the droplet velocity appears in the far-field and wall velocity condition. The droplet velocity is obtained by imposing an integral constraint following \cite{balestra2018viscous, mao2023effect}. For spatial discretization, we use quadratic element for velocity and linear element for pressure and surfactant concentration. We use triangular mesh elements of very fine mesh refinement close to the droplet interface and the nearby boundary. For time discretization, we use 2nd order backward differentiation formula (BDF). In-built parallel direct sparse solver (PARDISO) is used to solve the discretized linear system with relative tolerance of $10^{-6}$. We have validated our model with the existing literature (see Appendix A2: Fig.~\ref{fgr:9}). 

\section{Results: Evolution of flow field}
\begin{figure*}[ht!]
 \centering
\includegraphics[width=16cm]{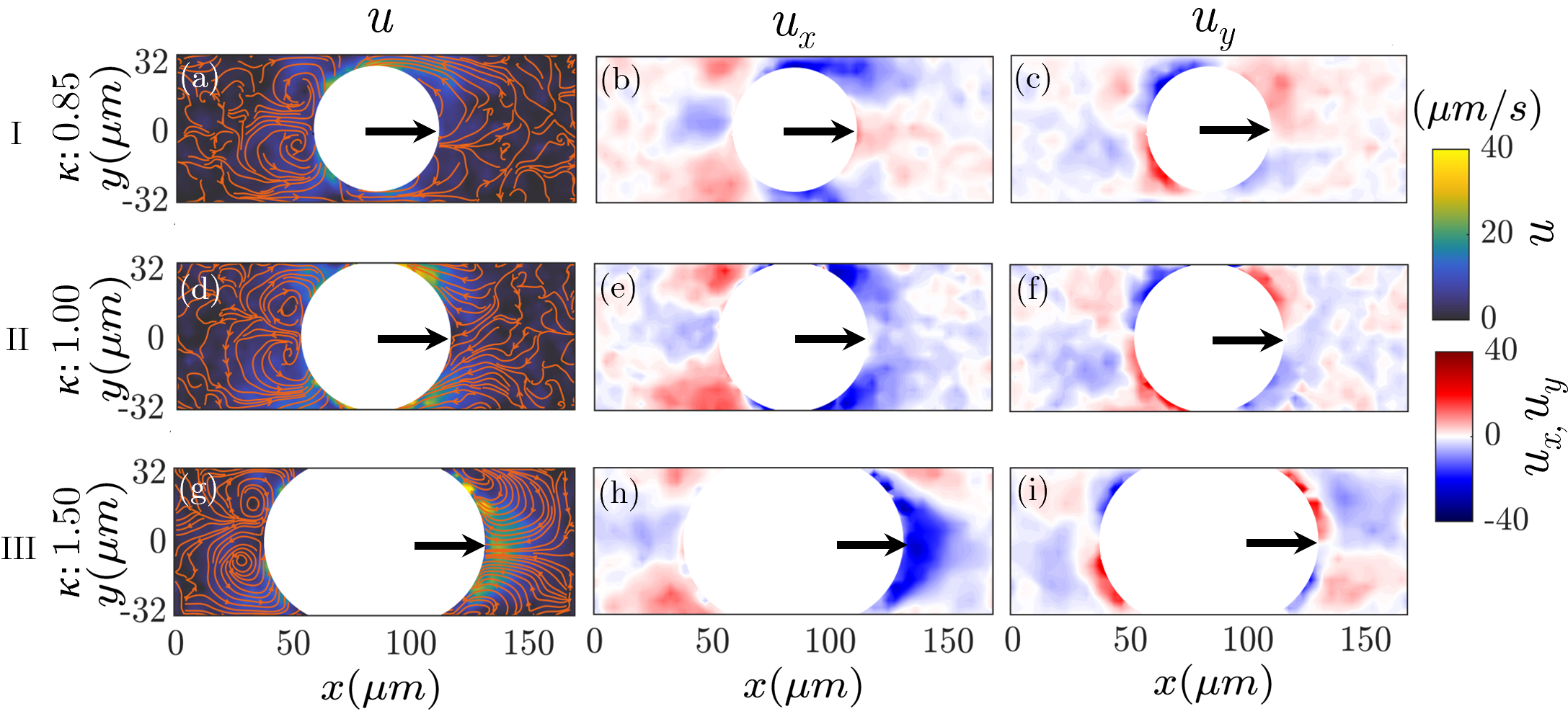}
\caption{\label{fgr:2}Variations in the velocity field for swimming active droplets with increasing effect of the confinement ($\kappa$) as determined using fluorescence microscopy and $\mu$PIV analysis-- \textbf{(a)-(c)} Regime-I ($\kappa \sim 0.85$), \textbf{(d)-(f)} Regime-II ($\kappa \sim 1.00$), and \textbf{(g)-(i)} Regime-III  ($\kappa \sim 1.50$). In the left column, \textbf{(a)}, \textbf{(d)}, and \textbf{(g)} depict contour plots of the resultant local velocity ($u$), along with the streamlines. The middle column: \textbf{(b)}, \textbf{(e)} and \textbf{(h)}, and the right column: \textbf{(c)}, \textbf{(f)} and \textbf{(i)}, represent the distributions of the axial $(u_{x})$ and transverse $(u_{y})$ components of the local disturbance velocity respectively. The black arrow denotes the direction of droplet motion.}
\end{figure*}

In a quasi-2D reservoir, the active droplets of sizes comparable to those used in the experiments swim with a steady velocity, and by generating a `weak-pusher' type velocity field (see SM-Fig.~\ref{fgr:11}). 
In confinements, the swimming velocity of isotropic, autophoretic spherical particles increases with increasing effect of the confinement, as given by $\kappa$, reaching a peak value at a specific threshold ($\sim 0.7-0.8$ for a spherical, isotropic autophoretic particle in a capillary \cite{francesco2022confined}).
However, in narrow or tight confinements, like in microchannels and microcapillaries with $\kappa > 0.7-0.8$, the swimming velocity of active droplets gradually decreases towards a constant value with increasing $\kappa$ \cite{de2021swimming}.
This is also observed for the experiments reported here (Fig.~\ref{fgr:1}(d)). 
In this regime, over which $u_d$ decreases with increasing $\kappa$, hydrodynamics in the thin lubrication film between the active droplet and the microchannel walls plays a dominant role in dictating the swimming dynamics \cite{francesco2022confined,de2021swimming}.
Here, we explain the evolution and characteristics of the hydrodynamic signature, or the flow field, generated by the active droplet as it swims in this same regime.
To this end, we divide the swimming of the active droplet with varying $\kappa$ into four regimes (Fig.~\ref{fgr:1}(d)) based on the observed characteristics of the velocity field.

\subsection*{Regime-I: $0.70$ $<\kappa<$ $0.90$}
\begin{figure*}
 \centering
 \includegraphics[width=16cm]{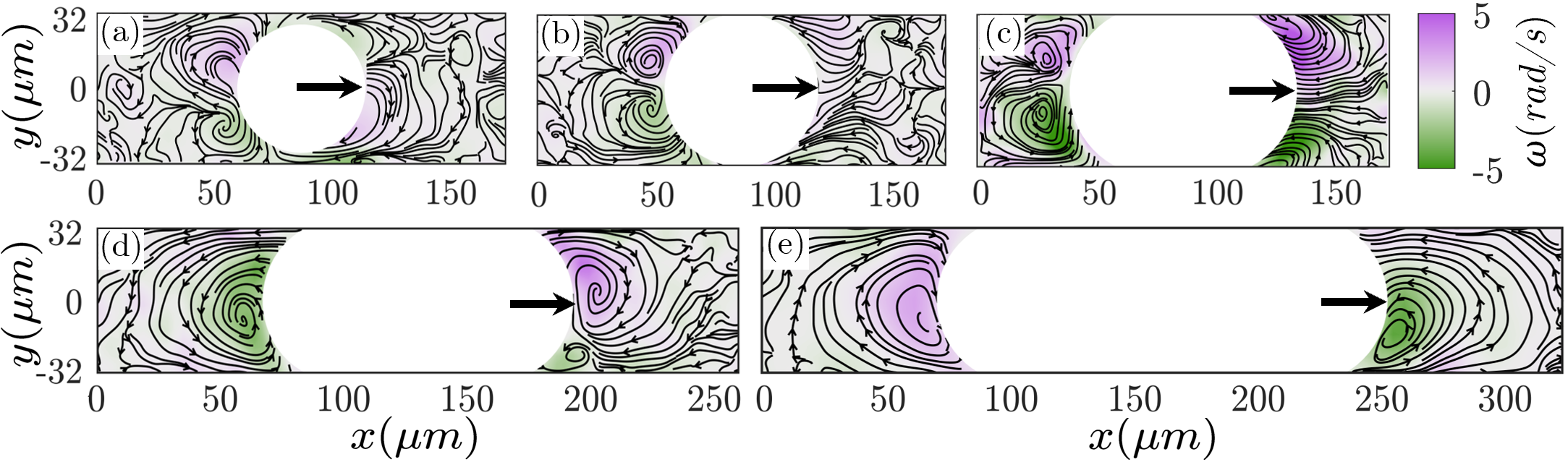}
 \caption{\label{fgr:4} Experimentally evaluated variations in the vorticity $(\omega)$ distribution for the active droplet velocity field with increasing effect of the confinement $(\kappa)$. From \textbf{(a)} to \textbf{(e)}, $\kappa$ increases as-- $\kappa=$ $0.85$, $\kappa=$ $1.00$, $\kappa=$ $1.50$, $\kappa=$ $2.25$, and $\kappa=$ $3.00$.}
\end{figure*}

In Regime-I, the spherical droplet microswimmer swims by generating a `pusher' type velocity field, i.e. by pushing liquid from the anterior and posterior ends with higher local velocity $(u)$ in the posterior interfacial region between the microswimmer and the microchannel walls (Fig.~\ref{fgr:2}(a); see Supplemental Video S1).
The generated velocity field is approximately symmetric about the x-axis, with two circulation zones in each of the anterior and posterior regions of the droplet microswimmer (Fig.~\ref{fgr:2}(a)).
The orientation of the vorticity $(\omega)$ associated with the two circulations zones (Fig.~\ref{fgr:4}(a)), both at the anterior and posterior, are opposite to each other.
Here, negative and positive values of $\omega$ represent anti-clockwise (orientation along positive z-axis) and clockwise (orientation along negative z-axis) orientations of the vorticity, or angular velocity, of the flow field about the positive x-axis.
The distributions of the axial component $(u_x)$ (Fig.~\ref{fgr:2}(b)) and the transverse component $(u_y)$ (Fig.~\ref{fgr:2}(c)) of the local velocity are approximately symmetric and skew-symmetric about the x-axis respectively.
$u_x$ about the anterior apex is comparable to the steady swimming velocity of the droplet microswimmer (Fig.~\ref{fgr:2}(b)).
Note that the disturbance velocity diminishes away from the droplet microswimmer as expected. 
In this regime, as the droplet microswimmer swims through the microchannel it leaves behind a trail of filled micelles which tend to accumulate close to the posterior apex (i.e. about an orientation angle of $\pi$ relative to the positive x-axis), with a slight displacement towards the adjacent wall (kymograph in Fig.~\ref{fgr:5}(a)).
The dark bands in the kymographs represent the microchannel walls.    

\begin{figure*} [ht!]
 \centering
 \includegraphics[width=16cm]{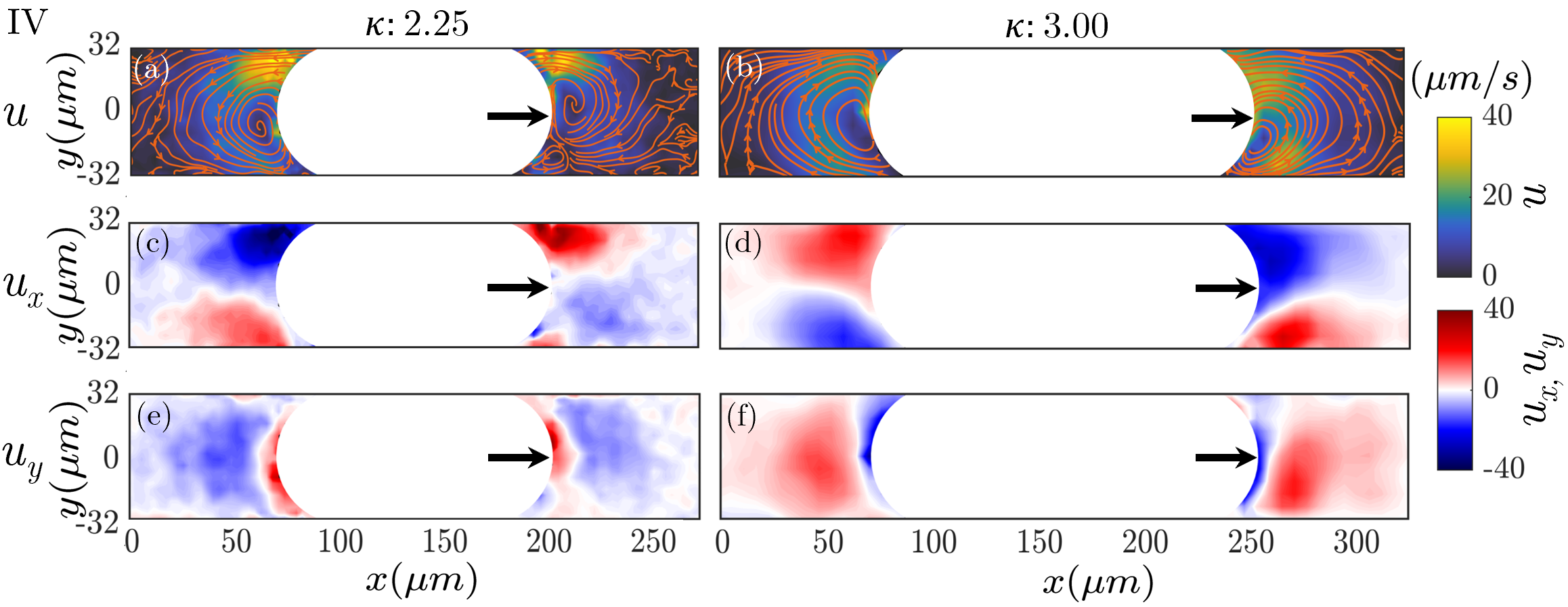}
 \caption{\label{fgr:3} Velocity fields for elongated (capsule-like) active droplets for $\kappa$ $>1.80$ (Regime-IV) evaluated using fluorescence microscopy and $\mu$-PIV analysis. The first row: \textbf{(a)} and \textbf{(b)}, represents the contour plots of the resultant local velocity ($u$), along with the streamlines. The second row: \textbf{(c)} and \textbf{(d)}, and the third row: \textbf{(e)} and \textbf{(f)}, respectively represent the corresponding distributions of the axial $(u_{x})$ and transverse $(u_{y})$ components of the local velocity.}
\end{figure*}

\subsection*{Regime-II: $0.90$ $<\kappa<$ $1.30$}
In Regime-II, the tight fitting, almost spherical droplet microswimmer switches to a `puller' type velocity field.
It swims by apparently pulling liquid at the front, and by pushing liquid at the back (Fig.~\ref{fgr:2}(d); see Supplemental Video S2). 
In this case, $u$ is relatively higher in the anterior, instead of the posterior, interfacial region between the microswimmer and the microchannel walls (compare Figs.~\ref{fgr:2}(d) and (a)). 
In this regime, the two circulation zones at the anterior of the droplet microswimmer become weaker, as highlighted by the corresponding smaller values of $\omega$ (Fig.~\ref{fgr:4}(b)). 
The circulation zones at the posterior of the droplet microswimmer still persist (Fig.~\ref{fgr:4}(b)) with $\omega$ comparable to that in Regime I.

\begin{figure*} [ht!]
 \centering
 \includegraphics[width=16cm]{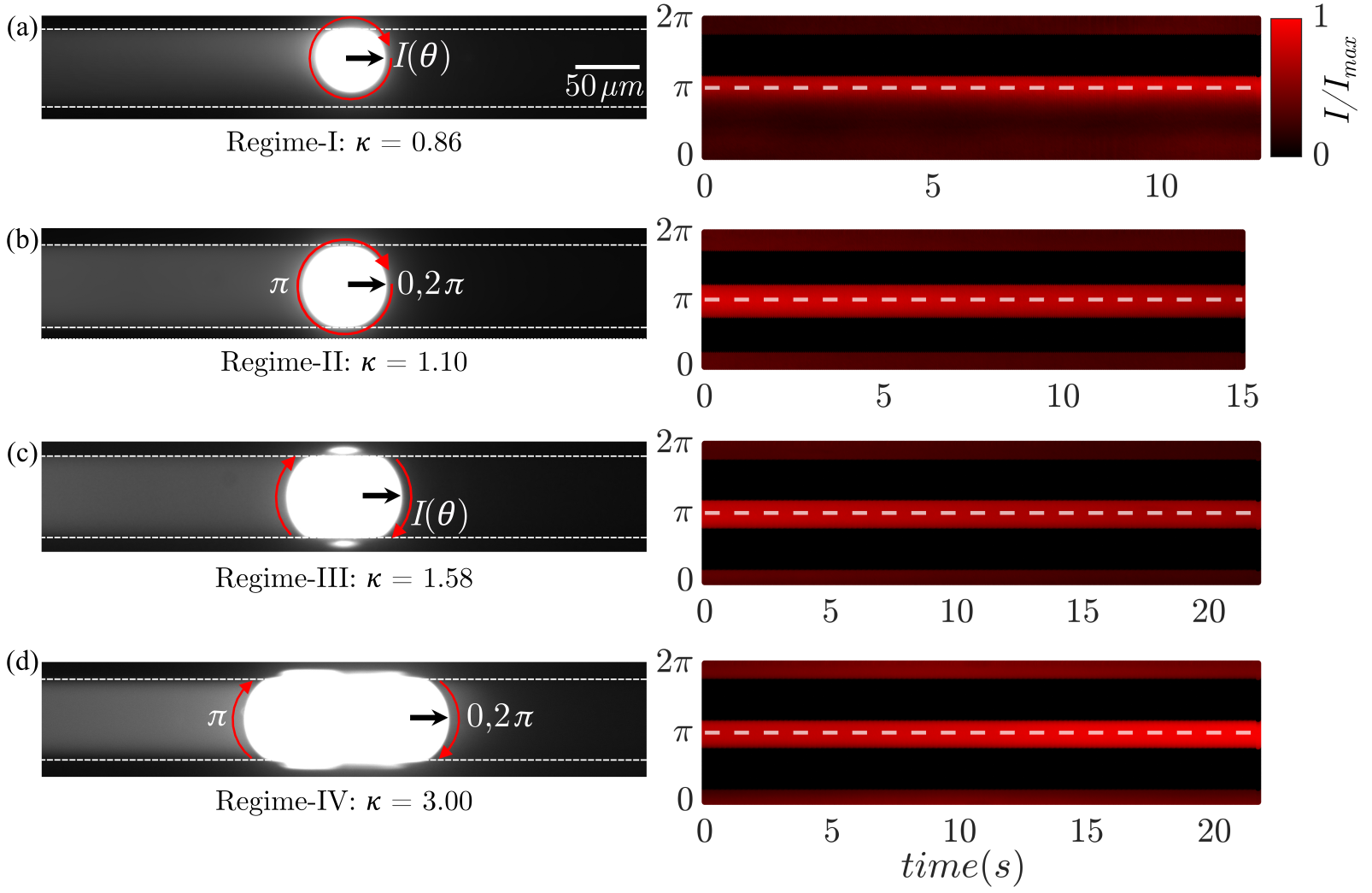}
 \caption{\label{fgr:5} Fluorescence microscopy images showing the filled micelle trail behind the active droplet (left column), and the corresponding kymographs (right column) representing the temporal evolution of the intensity profile $I(\theta)$ (filled micelle concentration) around the droplet interface over 20 s of propulsion. From top row to bottom row, results are presented for increasing effect of the confinement $(\kappa)$ over the four regimes. For the kymographs, the fluorescence intensity ($I$) for the filled micelle is normalized by the maximum interfacial intensity ($I_{max}$) for $\kappa=$ $3.00$.}
\end{figure*}

In Regime-II, the distribution of $u_x$ in the anterior region of the droplet microswimmer is apparently completely negative (Fig.~\ref{fgr:2}(e)). 
This seems contradictory to the fact that the droplet is still swimming along the positive x-axis, as has been also noted previously \cite{de2021swimming}.
However, using high resolution $\mu$PIV analysis, we observe that  $u_x$ is actually positive near the anterior apex, with values comparable to the droplet swimming velocity (see SM-Fig.~\ref{fgr:12}). 
Hence, in Regime-II, the velocity field of the droplet microswimmer is like that of a `puller', with the `pulling' signature evident from the immediate vicinity of the anterior apex of the microswimmer, instead of in the far-field as in classical pullers \cite{lauga2020fluid}.
This also implies that the stagnation point is imperceptibly close to the anterior apex of the droplet.
The distribution of $u_y$ remains skew-symmetric about the x-axis, and comparable to that in Regime-I (compare Figs.~\ref{fgr:2}(f) and (c)). 
In Regime-II, the filled micelle trail spans uniformly over the entire width of the microchannel at the posterior end of the droplet microswimmer, instead of about the posterior apex as in Regime-I (kymograph in Fig.~\ref{fgr:5}(b)).

\subsection*{Regime-III: $1.30$ $<\kappa<$ $1.80$}
In Regime-III, the droplet microswimmer adapts to the tighter confinement by assuming a stadium-like shape \cite{Stadium} (Fig.~\ref{fgr:2}(g)).
It still swims like a `puller' by pulling liquid at the front with a velocity field that is symmetric about the x-axis as in Regime-II (Fig.~\ref{fgr:2}(g); see Supplemental Video S3). 
But in this case, there are three significant differences -- (i) the higher value of $u$ is displaced more towards the anterior apex away from the interfacial region between the microchannel wall and the microswimmer as in Regime-II (compare Figs.~\ref{fgr:2}(g) and (d)); (ii) the two circulation zones at the anterior of the droplet microswimmer becomes stronger, as highlighted by the larger values of $\omega$ associated with these zones (compare Figs.~\ref{fgr:4}(c) and (b)); (iii) the distribution of $u_x$ becomes increasingly negative in the anterior region of the droplet microswimmer (compare Figs.~\ref{fgr:2}(h) and (e)).
In this case, we think that there is still an increasingly small region of positive $u_x$ about the anterior apex of the droplet microswimmer, which is hard to resolve even with our high resolution $\mu$PIV.

In Regime III, the two circulation zones at the posterior of the droplet microswimmer still persist, but with relatively larger values of $\omega$ as compared to Regime-II (compare Figs.~\ref{fgr:4}(c) and (b)).
The distribution of $u_y$ (Figs.~\ref{fgr:2}(i)), and the characteristics of the filled micelle tail at the posterior of the droplet microswimmer (kymograph in Fig.~\ref{fgr:5}(c)) remain qualitatively similar to those in Regime-II.

\subsection*{Regime-IV: $\kappa>$ $1.80$}
Finally, in Regime-IV, the droplet microswimmer swims in the tighter confinement by adopting an elongated capsule-like shape \cite{Capsule} (Fig.~\ref{fgr:3}(a), (b); See Supplemental Videos S4 and S5). Interestingly, in this regime the velocity field (flow pattern) becomes asymmetric about the x-axis (Fig.~\ref{fgr:3}(a), (b)).
This asymmetric swimming regime has been partially observed before in microcapillaries \cite{de2021swimming}. 
Here we explain in detail the salient features of this unique regime.\\

The gradual evolution of the asymmetric flow pattern from the symmetric one, i.e. the emergence of Regime IV from Regime III, with increasing confinement effect is shown in SM-Fig.~\ref{fgr:13}.
The flow pattern in Regime IV is characterized by two dominant circulation zones, one each at the anterior and posterior of the droplet microswimmer (Fig.~\ref{fgr:3}(a), (b)).
The orientation of $\omega$ corresponding to these two circulation zones are always opposite to each other, i.e. if $\omega$ for the anterior zone is clockwise then that for the posterior zone is always anti-clockwise, and vice versa (compare Figs.~\ref{fgr:4}(d), (e)).
Note that with the variation in $\kappa$ within Regime-IV, sometimes there is an emergence of a secondary weak circulation zone in the anterior of the droplet microswimmer (Figs.~\ref{fgr:3}(a) and~\ref{fgr:4}(d)).
This secondary circulation zone is unstable as its strength fluctuates with increasing $\kappa$ within Regime-IV (SM-Fig.~\ref{fgr:14}).
But when this secondary circulation zone does exist, $\omega$ associated with it is always opposite to that of the dominant circulation zone at the anterior (Fig.~\ref{fgr:4}(d) and SM-Fig.~\ref{fgr:14}(b), (c), and (e)).  

In Regime-IV, on one hand, the distribution of $u_x$ becomes skew-symmetric, instead of symmetric, about the x-axis (Fig.~\ref{fgr:3}(c), (d)), while on the other, the distribution of $u_y$ loses its skew-symmetry about the x-axis (Fig.~\ref{fgr:3}(e), (f)). 
In this regime also, the filled micelle trail covers the entire width of the microchannel. 
But the concentration of the filled micelle at the posterior of the droplet microswimmer is higher compared to the other regimes (Fig.~\ref{fgr:5}(d); compare the kymographs).
We also think that in this regime, there is non-zero concentration of filled micelles even in the anterior region of the microswimmer (compare the kymographs in Fig.~\ref{fgr:5}).  

\section{Discussions}
\begin{figure} [ht!]
 \centering
 \includegraphics[width=8cm]{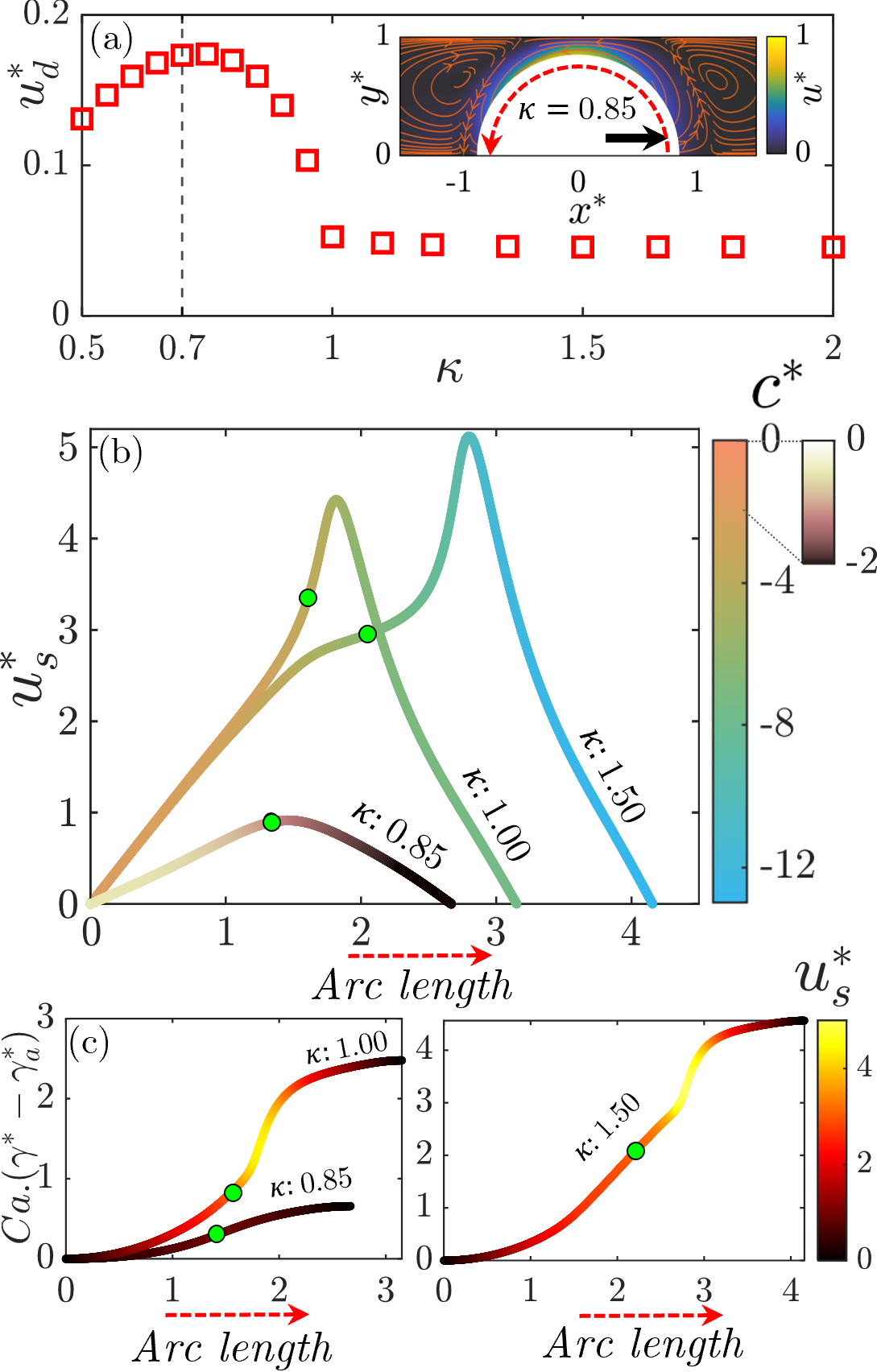}
 \caption{\label{fgr:6} Results from the numerical simulations. \textbf{(a)} Variation of the non-dimensional droplet swimming velocity $(u_d^*)$ with increasing confinement effect $(\kappa)$. The inset shows the velocity contour plot along with the streamlines for the droplet microswimmer in Regime-I. 
 \textbf{(b)} Variations of the non-dimensional droplet interfacial velocity $(u_s^*)$ with the arc length along the droplet interface for increasing values of $\kappa$ over Regimes I-III. The arc length increases from the anterior apex $(0)$ towards the posterior apex (dotted red arrow in inset in (a); also see Fig.~\ref{fgr:7}). The graphs are colour-coded with the non-dimensional surfactant monomer concentration $(C^*)$ along the interface.
 \textbf{(c)} Variations of the non-dimensional interfacial tension $(\gamma^*)$ for the active droplet for Regime-I: $\kappa$ $=0.85$, Regime-II: $\kappa$ $=1.00$ (left), and Regime-III: $\kappa$ $=1.50$ (right). $\gamma^*$ is rescaled to the interfacial tension at the anterior apex $(\gamma^*_a)$, and then multiplied by the constant capillary number $(Ca)$.The graphs are colour-coded with the corresponding magnitude of $u_s^*$.  
 The green circles represent the location of the equator for spherical, or almost spherical, droplets, and the location of the middle transverse plane for the stadium-shaped droplet. We consider the following simulation parameters: $Re=10^{-3}$, $Ca=10^{-3}$, $\rho_i/\rho_e=1$, $\mu_i/\mu_e = 138$ and $Pe = 10$.}
\end{figure}

\begin{figure}[ht!]
 \centering
 \includegraphics[width=8.5cm]{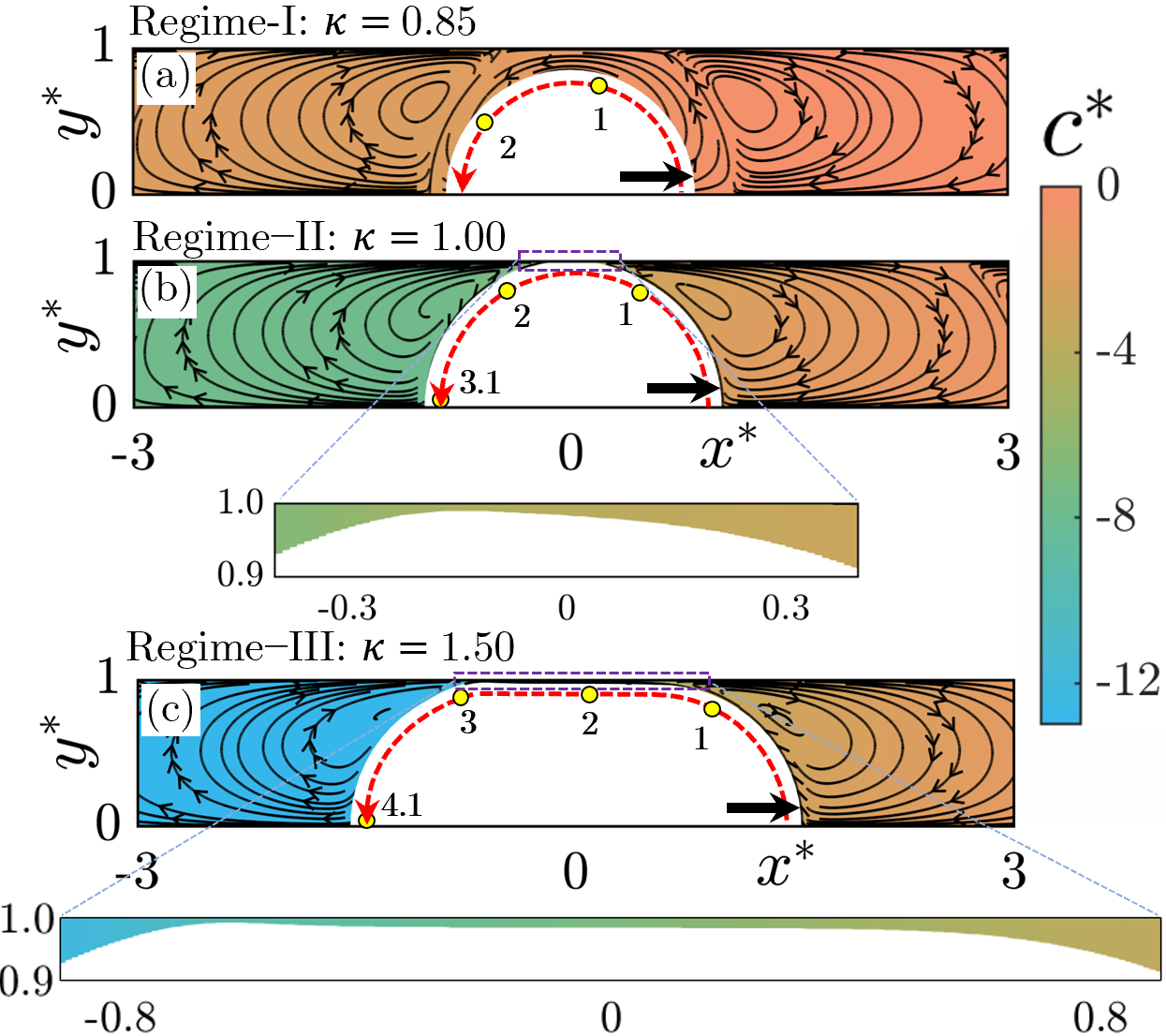}
 \caption{\label{fgr:7} Results from numerical simulation. \textbf{(a)}, \textbf{(b)} and \textbf{(c)} represent the streamlines for the velocity field generated by the active droplet for Regime-I: $\kappa=$ $0.85$, Regime-II: $\kappa=$ $1.00$, and Regime-III: $\kappa=$ $1.50$, respectively. The streamlines are plotted on top of the corresponding contour plots for the non-dimensional surfactant monomer concentration $(C^*)$. For the axisymmetric simulations, only the top half has been shown here.}
\end{figure}

We explain the aforementioned evolution of the velocity field generated by the active droplet as it squeezes through increasingly tighter confinement (increasing $\kappa$) using the numerical simulation results shown in Figs.~\ref{fgr:6} and~\ref{fgr:7}.
The numerical simulation captures the gradual increase in the non-dimensional swimming velocity $(u_d^*)$ with increasing $\kappa$ for weak confinements $(0.50<\kappa<0.70)$ (Fig.~\ref{fgr:6}(a)), as explained for isotropic autophoretic particles \cite{francesco2022confined}. 
Importantly, the simulation also captures the gradual decrease in  $u_d^*$ with $\kappa$ $ (>0.70)$, as observed in the experiments (Fig.~\ref{fgr:1}(d)). 

In Regime-I $(\kappa \sim 0.85)$, the non-dimensional concentration of the surfactant monomer $(C^*)$ reduces along the interface of the spherical active droplet, away from the anterior apex (Fig.~\ref{fgr:6}(b)). 
This is due to the spontaneous symmetry breaking of the solute concentration under dominant advective effect, accentuated by the confinement \cite{francesco2022confined}.  
Accordingly, the interfacial surface tension $(\gamma^*)$ gradually increases away from the anterior apex of the droplet microswimmer (Fig.~\ref{fgr:6}(c)).
This results in the gradual increase of the Marangoni-stress driven interfacial velocity $(u^*_s)$ (Fig.~\ref{fgr:6}(b)), responsible for the self-propulsion of active droplets. 
Note that the peak value of $u^*_s$ is reached in the posterior half of the droplet microswimmer, just beyond the equator (green circle in Fig.~\ref{fgr:6}(b)). 
This variation of $u_s^*$ is commensurate with the `pusher' type velocity field observed in Regime-I (compare inset in Fig.~\ref{fgr:6}(a) and Fig.~\ref{fgr:2}(a); also compare streamlines in Figs.~\ref{fgr:7}(a) and~\ref{fgr:2}(a)). 

In Regime-II $(\kappa \sim 1.00)$, the gradient (decrease) in $C^*$ along the droplet interface is relatively higher over the anterior half of the tight fitting, almost spherical droplet microswimmer (Fig.~\ref{fgr:6}(b)).
Consequently, there is steeper increases in $\gamma^*$ over the anterior half (Fig.~\ref{fgr:6}(c)).
This results in almost two-three times increase in $u_s^*$ over the anterior half of the droplet microswimmer compared to Regime-I (Fig.~\ref{fgr:6}(b)). 
This enhanced $u_s^*$ is responsible for driving the liquid through the thin lubrication film which now exists between the droplet interface and the microchannel wall, about the droplet equator. 
For the droplet microswimmer to swim through the tight microchannel, it must transport the displaced liquid from the anterior to the posterior through the thin liquid film, which is facilitated by the higher value of $u_s^*$. 
The `puller' type velocity field observed in Regime-II (compare streamlines in Figs.~\ref{fgr:7}(b) and~\ref{fgr:2}(d)) emerges out of this necessity of transporting liquid through the thin film, as driven by the higher $u_s^*$ in the anterior part of the droplet microswimmer. 
The distributions of $u_x$, $u_y$, and $\omega$ follow consequently. 

Note that the increase in $u_s^*$ does not translate to increased $u_d^*$ in Regime-II.
This can be understood as follows -- a balance between the energies associated with the motion of the droplet microswimmer and the viscous dissipation in the thin film region shows that $u_d^* \sim u_s^* (1-\kappa)^{1/2}$.
From Regime-I to Regime-II, $(1-\kappa)$ reduces from $\sim O(10^{-1})$ to at least $\sim O(10^{-3})$, assuming that the thickness of the intervening thin film in Regime-II is $\sim O(100)$ nm, if not smaller.
Hence, $u_d^*$ reduces since $(1-\kappa)$ reduces by two orders of magnitude, while $u_s^*$ still remains $O(1)$ (Fig.~\ref{fgr:6}(b)).

The peak in $u_s^*$ beyond the equator observed in the numerical simulations (Fig.~\ref{fgr:6}(b)) is due to a local sharp reduction in $C^*$ (and hence, a sharp increase in $\gamma^*$; Fig.~\ref{fgr:6}(c)) stemming from a local deformation in the droplet interface towards the posterior end of the thin film region (Fig.~\ref{fgr:7}(b)). 
We are unable to clearly discern such deformation in the experiments reported here.
Beyond the location of this peak in $u_s^*$, the gradient in $C^*$ becomes weak.
$C^*$ asymptotically approaches an almost constant value over the posterior droplet interface (Fig.~\ref{fgr:6}(b)).
In the thin lubrication film between the droplet interface and the non-penetrable, rigid wall, the concentration of surfactant monomers tends to rapidly homogenized due to the faster diffusion-driven transport in the thin region.
Given that the surfactant consumption rate is also constant over the entire droplet interface, there is stronger depletion of $C^*$ (Fig.~\ref{fgr:7}(b)) over the thin film region (steeper increase in $\gamma^*$; Fig.~\ref{fgr:6}(b)), especially towards its posterior end. 
Beyond this, once the thin region starts widening, the surfactant monomer concentration is already strongly depleted, and this results in a weak gradient in $C^*$ over the posterior droplet interface. 
Accordingly, $\gamma^*$ becomes asymptotically constant over the posterior interface (Fig.~\ref{fgr:6}(c)), beyond the peak in $u_s^*$, resulting in a sharp decrease in $u_s^*$.  
Hence, the pushing of the liquid at the posterior end, with two circulation zones, is not a Marangoni-stress dominated phenomenon, but also a consequence of the continuity of the flow between the thin film region and the posterior part of the droplet microswimmer.
Overall, there appears to be a sharp fore-aft contrast in $C^*$ over the droplet microswimmer length scale (i.e. $D_d$) in Regime-II (contour plot in Fig.~\ref{fgr:7}(b)).
But the majority of the reduction in $C^*$ is actually over the anterior part of the microswimmer and the thin film region, and not over the posterior part (Fig.~\ref{fgr:6}(b)). 
The sharp fore-aft contrast in the filled micelle concentration observed across the droplet microswimmer from the kymograph for Regime-II (Fig.~\ref{fgr:5}(b)) is analogous to the aforementioned distribution of $C^*$. 
The kymograph also establishes the stability of the physico-chemical mechanism underlying the droplet microswimmer propulsion in Regime-II.

In Regime-III $(\kappa \sim 1.50)$, as the droplet microswimmer adapts to the tighter confinement by adopting a stadium-like shape, the thin film region gets extended (Fig.~\ref{fgr:7}(c)).
Hence, the elongated droplet microswimmer maintains its `puller'-like velocity field to transport liquid from the anterior to the posterior through the extended lubrication film (compare streamlines in Figs.~\ref{fgr:7}(c) and~\ref{fgr:2}(g)).
The shape of the droplet microswimmer necessitates the greater magnitude of $\omega$ for the anterior circulation zones.
Over the ensuing thin film region, the gradient (decrease) of $C^*$ along the droplet interface is almost linear culminating in a corresponding linear increase in $\gamma^*$ (Fig.~\ref{fgr:6}(c)). 
This results in an almost uniform value of $u_s^*$ over the thin film region in Regime-III (Fig.~\ref{fgr:6}(b)).
Hence, although the variation in $u_s^*$ over the anterior droplet interface remains similar in Regimes II and III, the gradient of $u_s^*$ over the droplet interface in the thin film region becomes weaker in the latter (Fig.~\ref{fgr:6}(b)).
Accordingly, $u_s^*$ attains a smaller value in the thin film region for Regime-III as compared to Regime-II.
Consequently, $u_d^*$ is relatively smaller in  Regime-III, albeit slightly, following the aforementioned scaling argument, and assuming that $(1-\kappa)$ remains comparable in both the regimes. 
In Regime-III, there is more depletion of $C^*$ over the longer thin film region resulting in greater fore-aft surfactant monomer contrast compared to Regime-II (compare Figs.~\ref{fgr:7}(c) and (b)).   
It is difficult to comment whether this translates to greater concentration of filled micelles at the posterior of the droplet microswimmer in Regime-III, since the fluorescence intensity from the kymographs looks comparable for Regimes II and III (compare Figs. \ref{fgr:5}(b) and (c)).

It is impossible to capture the asymmetric velocity field for the droplet microswimmer in Regime-IV using the present axi-symmetric numerical simulation framework.
Furthermore, a direct 3D numerical simulation of the capsule-shaped droplet microswimmer is beyond the scope of the present work.
Based on our experimental observations, we speculate that as the droplet gets elongated, the droplet interface, and hence, the surrounding lubrication film, gets deformed in an asymmetric manner. 
The asymmetric velocity field is a consequence of the resulting asymmetric distribution of the interfacial velocity and the continuity of flow through the asymmetric thin film region. 
The extended thin film region and the single circulation zone at the posterior are responsible for the enhanced accumulation of the filled micelles as seen from the kymograph (Figs. \ref{fgr:5}(d)). 
Furthermore, the single circulation zone at the anterior may be responsible for a slight accumulation of filled micelles even at the front of the droplet microswimmer.

\section{Conclusions and outlook}
In this work, we explain the evolution in the velocity field generated by active droplets as they adapt their shape and autonomously squeeze through increasingly tighter microconfinements. 
We do this by first analyzing the alterations in the velocity field using fluorescence microscopy and $\mu$-PIV techniques.
Then we try to correlate the observed flow fields to the underlying physico-chemical hydrodynamics of active droplets in tight confinements, in the presence of a thin lubrication film.
This is achieved using the results from finite-element based numerical simulations.

In a strong confinement (Regime-I), the active droplet swims with a hydrodynamic signature that can be classified as a `pusher’ type velocity field. 
When the droplet microswimmer diameter becomes comparable to the microchannel length scale (Regime-II), for the still spherical droplet, the greater decrease in the concentration of the surfactant monomer over the anterior interface of the microswimmer results in stronger Marangoni-stress driven interfacial velocity.
This generates the `puller’ like hydrodynamic signature necessary to transport liquid through the thin film region between the tight fitting microswimmer and the microchannel walls. 
The dissipation in the thin lubrication film dictates that the terminal velocity of the microswimmer reduces despite the increase in the interfacial velocity of the microswimmer.
Furthermore, the greater decrease in the surfactant monomer concentration over the anterior part of the droplet microswimmer, specifically in the thin film region, results in sharper fore-aft concentration gradient across the droplet microswimmer as compared to Regime-I.
This is also evidenced from the filled micelle kymographs plotted for the active droplets.

When the microchannel length scale becomes smaller than the droplet microswimmer length scale, i.e. for confinement ratio greater than 1, the active droplet first adopts a stadium-like shape as it squeezes through the tighter confinement (Regime-III). 
As the thin lubrication film gets extended because of this shape, the microswimmer has to maintain its puller like velocity field in order to squeeze through the tight confinement by transporting liquid through the thin film.
Interestingly, the decrease in the interfacial surfactant concentration, and the corresponding variation in the interfacial tension, are almost linear in the extended thin film region.
Accordingly, the Marangoni stress-driven interfacial velocity is almost constant over the droplet interface in the thin film region culminating in the slight decrease in the terminal velocity in Regime III as compared to Regime II.
In Regime III, the extent of decrease in the surfactant concentration over the extended lubrication film is more, resulting in even sharper fore-aft gradient in the surfactant concentration across the droplet microswimmer as compared to Regime II.
Unfortunately, this cannot be clearly inferred by comparing the kymographs for Regimes II and III.   

As the confinement ratio further increases, the droplet microswimmer adopts an elongated, capsule-like shape (Regime-IV).
As the droplet microswimmer gets elongated, with increasing confinement, we speculate that non-axisymmetric modes get excited resulting in asymmetric shape of the thin lubrication film surrounding the microswimmer. 
The asymmetric shape of the thin film region may result in the asymmetric velocity field generated by the capsule-like droplet microswimmer. 
We further conclude that the single circulation zones observed in the anterior and posterior of the elongated droplet microswimmer are responsible for the enhanced concentration of filled micelles in the immediate vicinity of microswimmer, as observed in the corresponding kymographs.

In future, it will be interesting to investigate the extent to which the evolution in the hydrodynamic signature demonstrated by active droplets in increasingly tighter confinements- from pusher, to puller and to an asymmetric velocity field, is comparable to the hydrodynamic adaptation witnessed for biological microswimmers in tight confinements.
Such studies, focused on the characteristics of the velocity field generated by the microorganisms in increasingly tight confinements, can tell us whether the alteration in the nature of the flow field, as demonstrated here for self-propelled artificial microswimmers, plays a crucial role in the survival strategies of severely confined microswimmers by increasing their nutrient uptake.
The present results also indicate that since the velocity field generated by the self-propelled microswimmers changes, as their shape changes with increasing confinement, their mutual interactions are also bound to change with tighter confinements. 
Our work also provides an understanding of the chemo-hydrodynamic characteristics that can be present in possible state-of-the-art applications involving active droplets as autonomous cargo-carriers to severely narrow or constricted spaces, unreachable by passive carriers or even by catalytic active colloids.

 \section*{Acknowledgements}
SG and SSS acknowledge the Prime Minister's Research Fellowship (PMRF), a scheme by the Govt. of India to improve the quality of research in various research institutions in the country. SG and SSS thank Lakshmana Dora Chandrala for the discussions regarding the uncertainty calculations. SG and SSS acknowledge the help from Aneesha Kajampady in carrying out the pressure-driven flow experiments in the microchannel. 
SM gratefully acknowledges support from DST-SERB grant no. EEQ/2021/000561. SM thanks Indian Science Technology and Engineering Facilities Map (I-STEM), a Program supported by the Office of the Principal Scientific Adviser to the Govt. of India, for enabling access to the COMSOL Multiphysics 6.0 software used to carry out the numerical simulations.
RD gratefully acknowledges support from Science and Engineering Research Board (SERB), Department of Science and Technology (DST), Government of India, through grant no. SRG/2021/000892. RD also acknowledges support from IIT Hyderabad through seed grant no. SG 93. 

\appendix
\subsection*{Appendix A1: Details of the $\mu$PIV analysis}
The prominent sources of errors in the $\mu$PIV analysis arise from two important factors: the selection of the size and the properties of the tracer particles. The tracer particles are assumed to be neutrally buoyant and follow the flow streamlines when the difference between the fluid and tracer particle velocity is minimal. This can be achieved by reducing the size of the particles and reducing the difference in the density of the particles and the flow medium \cite{raffel2018particle}. 
To characterize the behaviour of the tracer particles in the flow medium, we determine the response time of the tracer particles towards a flow using Newton's second law as $t_{0} = \rho_{p} d_{p}^{2}/18 \mu_{f}$, where $\rho_{p}$ is the density and $d_{p}$ is the diameter of the tracer particles, $\mu_{f}$ is the dynamic viscosity of the fluid medium \cite{raffel2018particle}. The ratio of the response time of the particles to the advection time scale of the flow $l/u_{0}$ is known as Stokes number, where $l$ is the characteristic length scale which can be taken as $d_{p}$ and $u_{0}$ is the flow velocity.
The Stokes number should be less than 0.1 to maintain the errors due to tracers less than $1$$\%$ \cite{tropea2007springer}. The stokes number for the fluorescent particles of size $500$ $nm$ ($480\pm16$ $nm$ ) and density of $1055$ $kg/m^{3}$ is of the order $10^{-7}$. The velocity lag between the flow medium ($7.5\%$ aqueous surfactant solution) and the tracer particles is calculated using Stokes drag law given as $d_{p}^{2}(\rho_{p}-\rho_{f})g/18\mu$, where $g$ is acceleration due to gravity and subscripts $p$ and $f$ are for particle and fluid properties \cite{raffel2018particle}. The velocity lag for these tracer particles is of the order $10^{-2}$ $\mu m/s$. Hence, all the $\mu$PIV analysis has been carried out with these tracer particles to minimize the uncertainty due to the particle selection. 

These fluorescent particles are excited with a high-pressure mercury lamp (Nikon, Model$:$ LH-M100C-1,100W) through a dual-band filter with maximum excitation/emission wavelengths $487/562$ and 523/630 (Chroma Technology, Model$:$ 59009 - ET - FITC/CY3). These fluorescent particles are suspended in distilled water. We mix these suspended particles with $7.5\%$ aqueous surfactant solution in $1:20$ volume ratio and sonicate the solution for $10$ minutes to ensure proper mixing, and there is no coagulation. We then add the active droplets (stored in $0.1\%$ aqueous surfactant solution) to the sonicated $7.5\%$ surfactant solution containing the tracer particles (with the same ratio of $1:2$ as mentioned earlier) to visualize the disturbance field around the droplet swimmers after injecting in the microchannels.

The $\mu$PIV analysis is carried out using MATLAB-based software PIVlab. For the $\mu$PIV analysis using PIVlab, we use the Discrete Fourier Transform (DFT) method to calculate the cross-correlation matrix, which has the flexibility to give multiple passes. For the first pass, we take the $64 \times 64$ pixels interrogation area with a step size of $32 \times 32$ pixels followed by a second pass of $32 \times 32$ pixels interrogation area with a step size of $16 \times 16$ pixels with Gaussian 2x3-point sub-pixel estimator along with extreme correlation robustness.  

The accuracy of a digital PIV experiment is largely affected by errors in the measurement of the displacement, such as bias error ($\epsilon_{bias}$), which accounts for the accuracy of the measurements, and random error ($\epsilon_{rms}$) represents the precision in the measurements \cite{gui2002correlation}. Due to the uncertainty in the displacement measurement, errors are bound to arise in the velocity measurements. The velocity due to the displacement of the particles can be written as $v = s\Delta X/\Delta t$, where $s$ is the calibration factor, $\Delta X$ is the displacement of the particles in two consecutive frames, and $\Delta t$ is the time step in such consecutive frames. The uncertainty in the velocity measurement is calculated using $U_{v}=[(\frac{\Delta X}{\Delta t}U_{s})^{2}+(\frac{s}{\Delta t}U_{\Delta X})^{2}+(\frac{s \Delta X}{\Delta t^{2}}U_{\Delta t})^{2}]^{1/2}$ \cite{chandrala2016unsteady}, where $U_{s}$ is the uncertainty in calibration factor, $U_{\Delta X}$ is the total uncertainty in displacement measurement and $U_{\Delta t}$ is the uncertainty in time. In the present study, the time $\Delta t$ between two consecutive frames is $0.04$ $s$, and the uncertainty in time $U_{\Delta t}$ is $2\times 10^{-6}$ $s$. The calibration factor $s$ for a $30\times$ objective is $0.1762$ $\mu m$/pixel, and the uncertainty in the measurement of the calibration factor $U_{s}$ is $0.001379$ $\mu m$/pixel. The maximum displacement $\Delta X$ for a velocity of $40$ $\mu m/s$ (Fig.~\ref{fgr:2}) is taken as $9$ pixels. The uncertainty in displacement measurement $U_{\Delta X} =(\epsilon_{bias}^2+\epsilon_{rms}^2)^{1/2}$ for a particle image of $3\times3$ pixels (with a $30\times$ objective to resolve $500$ $nm$  particle diameter) using a DFT multipass algorithm is taken as $0.02$ pixel from \cite{thielicke2014flapping}. Using the above equation, the uncertainty in the velocity measurement is found to be $\pm 0.3225$ $\mu m/s$, which is $\pm 0.8$ $\%$ of the maximum velocity. 

\begin{figure}
 \centering
 \includegraphics[width=0.49\textwidth]{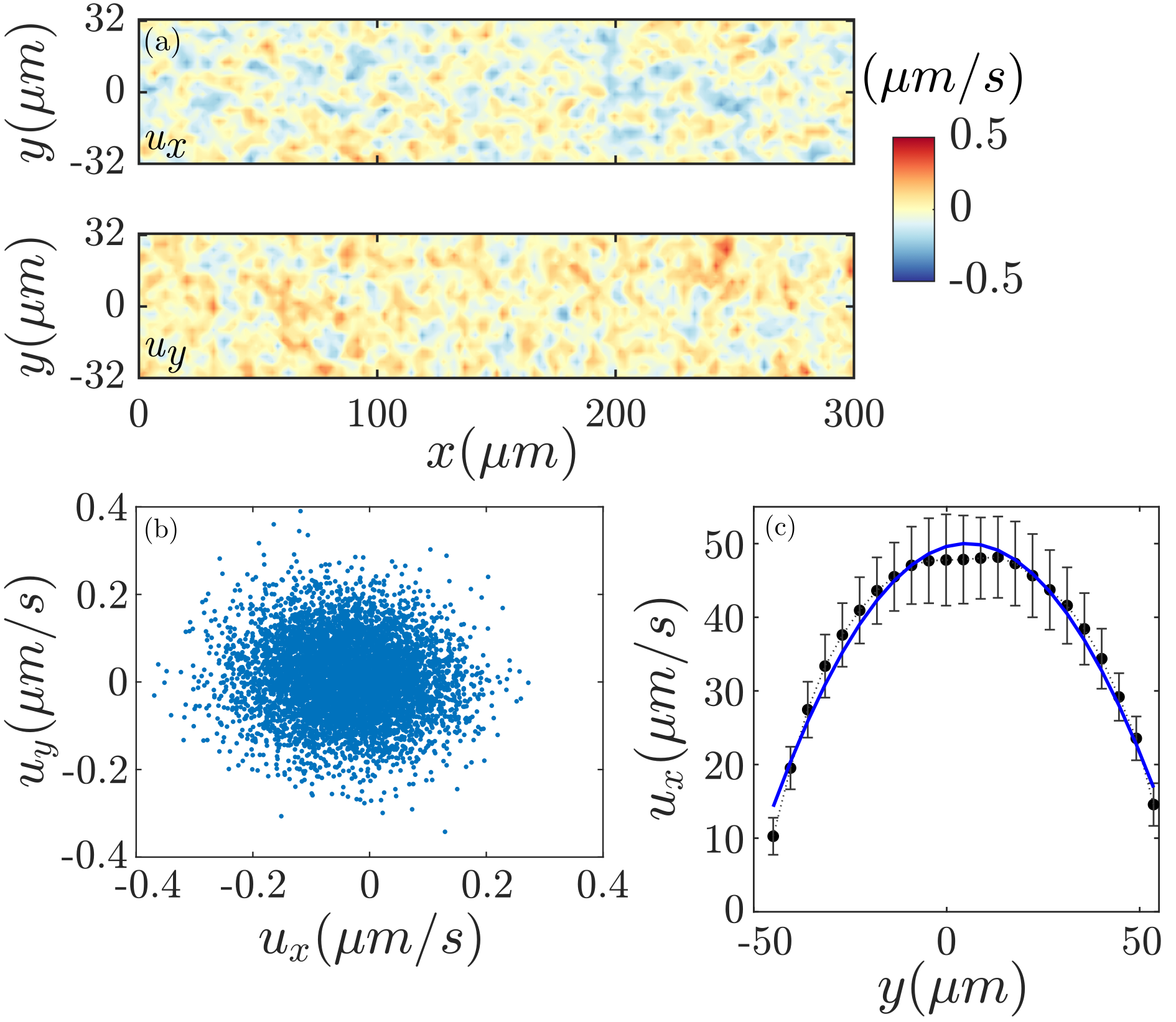}
 \caption{\label{fgr:8}$\mu$PIV analysis of quiescent medium seeded with $500$ $nm$ fluorescent particles exhibiting Brownian motion in a microchannel and with a prescribed pressure-driven flow. \textbf{(a)} The contour of $u_x$ and $u_y$ components of the particles within the microchannel. \textbf{(b)} Scatter plot of $u_x$ vs $u_y$ for the same. \textbf{(c)} Velocity measured using $\mu$PIV setup in a microchannel with the prescribed pressure-driven flow, a quadratic curve fitted to these data points along the channel width (solid blue line). $x$ is along the length, and $y$ is along the width of the microchannel as shown in Fig.~\ref{fgr:1}(a).}
 \end{figure}

We verify the results (flow field around the swimmer) obtained from the $\mu$PIV analysis are from the disturbance field generated due to the droplet self-propulsion and are not any noise coming from the analysis. To quantify the noise from the external medium in terms of velocity, experiments were carried out in a microchannel with a quiescent medium seeded with $500$ $nm$ fluorescent particles exhibiting Brownian motion.
The net displacement of the particles undergoing Brownian motion is close to zero over time, so we have calculated the velocities over a time span of $19$ $s$. The contour of the averaged x ($u_{x}$) and y ($u_{y}$) component's velocity for Brownian motion as shown in (Fig.~\ref{fgr:8}(a)). 
The minimum/maximum values of $u_{x}$ and $u_{y}$ components of velocity that we can detect for a particle undergoing Brownian motion in a quiescent medium is $-0.3690/0.2720$ $\mu m/s$ and $-0.3422/
0.3903$ $\mu m/s$ (see Fig.~\ref{fgr:8}(b)).
However, the average disturbance field velocity around the droplet for $\kappa = 0.85$ (Fig.~\ref{fgr:2} (a)) at a distance $\sim R_{d}$ from its circumference is $7.169 \pm 0.3312 \mu m/s$.
The velocity due to Brownian motion is very small compared to the magnitude of the disturbance field around the droplet, which implies that the velocity signal at a distance $\sim R_{d}$ from the circumference of the droplet is purely due to advection (self-propulsion) and not due to diffusive motion of the tracer particles.

We have also compared the velocity of a pressure-driven flow in the center plane calculated using the same $\mu$PIV setup in the $x$-direction (along the length) in a microchannel of the cross-sectional area of $100\times50$ $\mu m^2$ to a second-order polynomial, approximating a Hagen-Poiseuille flow in a channel as shown in Fig.~\ref{fgr:8}(c). The  $\mu$PIV velocity matches well with the second-order polynomial with an average standard deviation of $\pm 4.59704565$ $\mu m/s$, indicating the PIV setup is robust enough to capture the accurate displacements of the tracer particles in the flow. 

\subsection*{Appendix A2: Validation of numerical method}
\begin{figure}[ht!]
 \centering
\includegraphics[width=8cm]{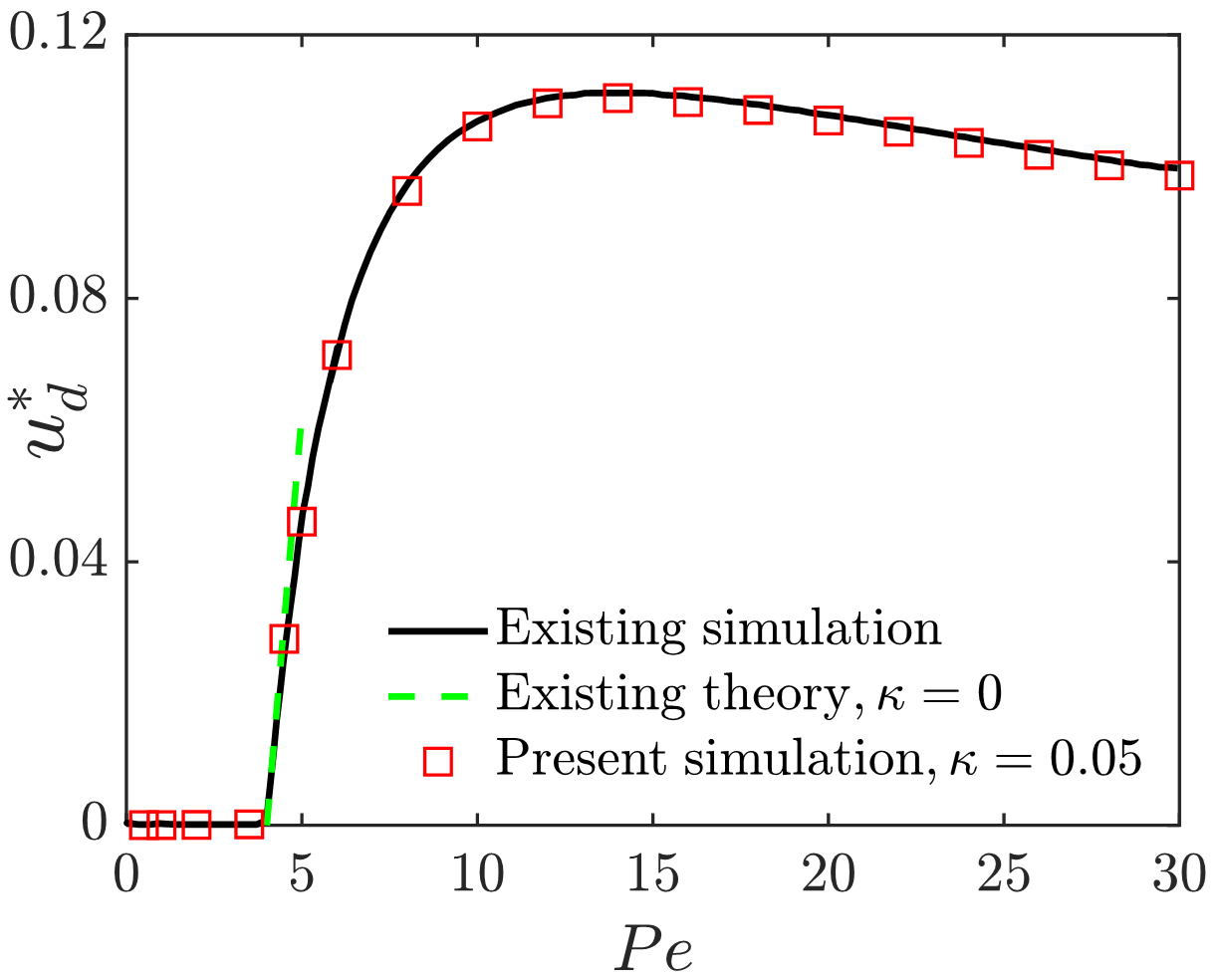}
 \caption{\label{fgr:9} Variation of steady swimming speed with P\'eclet number. The existing simulation data is taken from the supplementary material of \cite{izri2014self} and the theoretical expression is reported in \cite{morozov2019nonlinear}. We consider the following simulation parameters: $Re=10^{-3}$, $Ca=10^{-3}$, $\rho_i/\rho_e=1$, $\mu_i/\mu_e = 1/36$ and $\kappa = 0.05$. For this validation purpose, the droplet radius has been taken as the length scale.}
 \end{figure}

To assess the correctness of the present simulation method, we validate simulation results with existing literature. Fig.~\ref{fgr:9} shows the variation of steady swimming speed of unconfined droplet as a function of P\'eclet number. In the absence of bounding walls $(i.e.$ $\kappa=0)$, an active droplet remains motionless until $Pe=4$. At $Pe=4$ advection of surfactant transport is strong enough to make the motionless state unstable and the droplet starts to swim at a speed $(Pe - 4)/16$ near $Pe=4$ \cite{izri2014self, morozov2019nonlinear}. Fig.~\ref{fgr:9} shows that present simulation results compare well with the existing analytical and numerical results. To reduce the effect of bounding wall, we have chosen $\kappa=0.05$.      

\bibliography{active_lubrication}
\newpage
\clearpage

\title{Supplementary material for Flow fields around active droplets squeezing through tight confinements}
\maketitle
\author{Subhasish Guchhait}%
\affiliation{
 Department of Mechanical and Aerospace Engineering,\\
Indian Institute of Technology Hyderabad,
Kandi, Sangareddy 502285, India }
\author{Smita S. Sontakke}%
\affiliation{
 Department of Mechanical and Aerospace Engineering,\\
Indian Institute of Technology Hyderabad,
Kandi, Sangareddy 502285, India }
\author{Shubhadeep Mandal}%
\affiliation{
Department of Mechanical Engineering,\\
Indian Institute of Science,
Bengaluru 560012, India }
\author{Ranabir Dey}%
\affiliation{ 
Department of Mechanical and Aerospace Engineering,\\
Indian Institute of Technology Hyderabad,
Kandi, Sangareddy 502285, India }
%


\section*{I. Generation of active droplets}
\begin{figure*}
 \centering
\includegraphics[width=16cm]{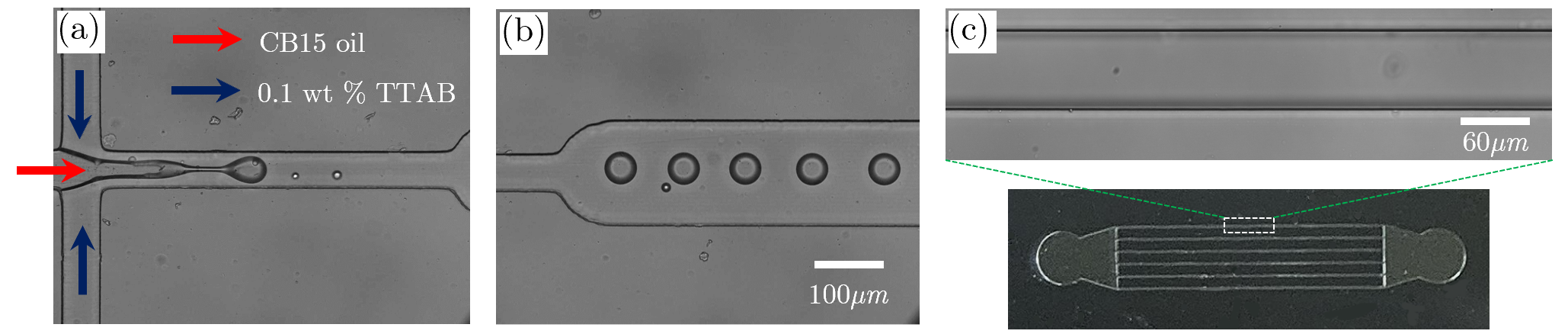}
 \caption{\label{fgr:10}\textbf{(a)} Generation of active droplets by the flow-focusing device. Here, the red and blue arrow indicates CB15 oil, 0.1 wt\% TTAB solution, respectively, \textbf{(b)} Formation of mono-disperse droplets, \textbf{(c)} Close-up view of the microfluidic chip with microchannels bonded to the coverslip.}
 \end{figure*}

We insert two syringe pumps in the flow-focusing device: one with 0.1 wt\% TTAB solution and another with CB 15 oil (Fig.~\ref{fgr:10}(a)). First, we fill out the whole channel with a TTAB solution-containing needle at a flow-rate of 500 $\mu l/hr$ so that the trapped air can go out quickly, and then we insert the oil needle and maintain its flow rate $\sim$ 5 $\mu l/hr$ while we slow down the TTAB flow rate to 100 $\mu l/hr$. Once the oil enters the inlet section, we reduce its flow rate to 1 $\mu l/hr$ and by tunning both the flow rates, we generate varying diameters of mono-disperse droplets (Fig.~\ref{fgr:10}(b)). After the generation, we switch off the oil flow-rate and increase the flow rate of the TTAB solution to approx 10000 $\mu l/hr$, and with the help of that flow, we preserve those droplets in a small container from the outlet, which is coated with plasma to prevent their adherence to the container's bottom. 

\section*{II. Droplet's centroid tracking method}
First, we crop the recorded video according to our region of interest. Next, we extract images from that cropped footage and convert them into binary images using in-house Matlab code. Once binarized, the image displays droplets and walls in white, with the rest represented in black (based on preference). The microchannel must undergo thorough cleaning to ensure the absence of any dust particles. Otherwise, it may also appear white, making droplet tracking challenging. When dealing with spherical droplets, we mark a solid red circle precisely around the white-coloured droplet in the initial figure. This enables the system to capture and retain the area. Subsequently, it autonomously scans for the remembered area in the succeeding images while simultaneously measuring the centroid. For elongated droplets, following binarization, we employ an in-house Matlab code to track the droplet-TTAB (7.5 wt $\%$) interface from both the left and right sides. Subsequently, utilizing these tracking points, the system computes the centres of the left and right circles, which constitute part of the interface's arc. After obtaining the center points, to determine the y-center of the elongated droplet, we calculate the mean of all the y-coordinates of the centers of the left and right circles (since the droplet can only move along the X-axis). For the x-center, we compute the mean of the x-coordinates of those circles in each image. Thus, we track the centroids at each time step (0.04 sec), and once we get those, we can calculate the instantaneous velocity of the droplet easily.

\begin{figure*}[h]
 \centering
 \includegraphics[width=0.7\textwidth]{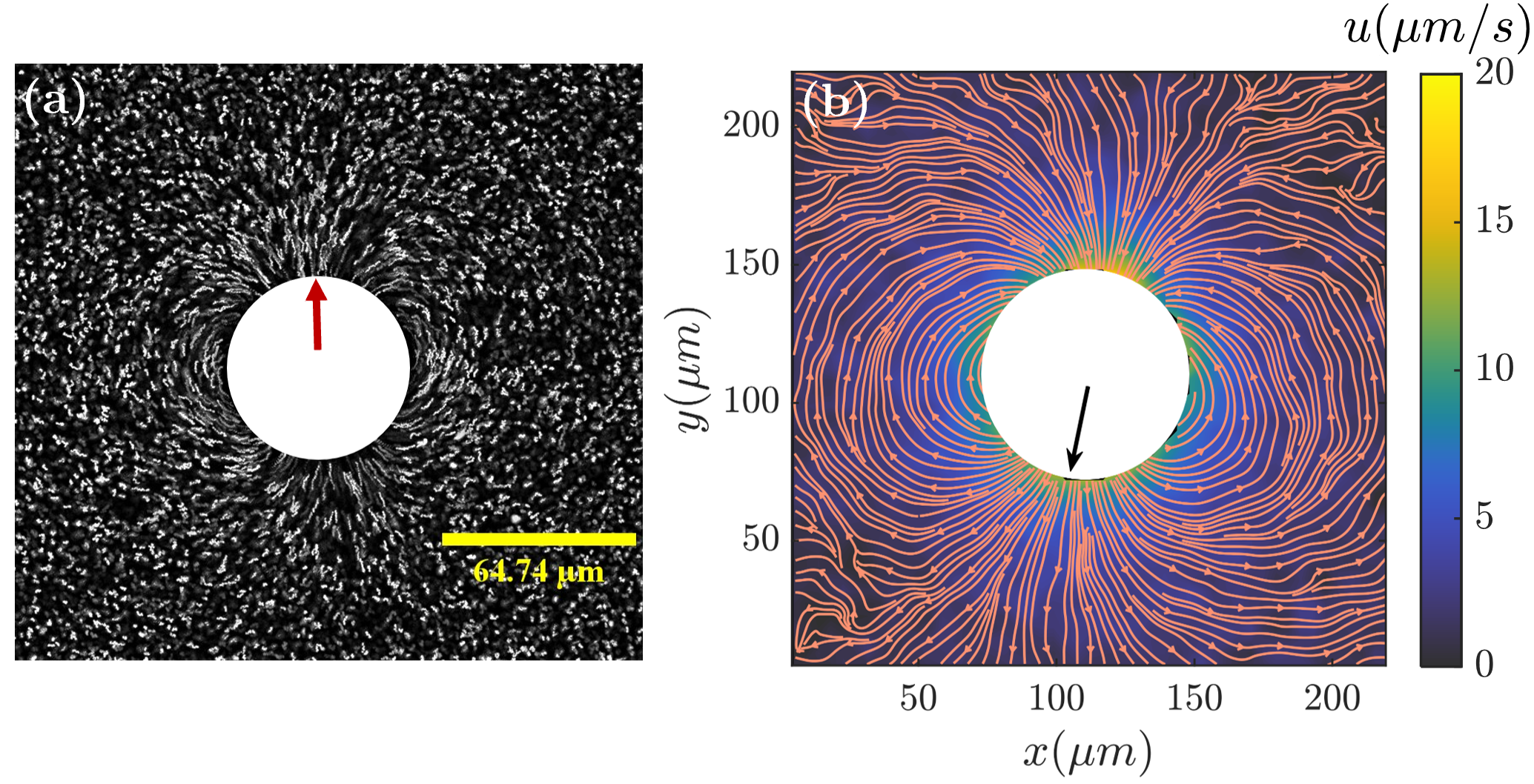}
 \caption{\label{fgr:11}The flow field of an active microswimmer within a 2D reservoir is depicted in the presented figures. \textbf{(a)} Here, we present the flow trace image, generated from the raw images of the tracer particles using ImageJ software \cite{schneider2012nih}.\textbf{(b)} The streamline plot generated by PIV analysis is represented. Based on this observation, it can be concluded that an active droplet exhibits a dipolar flow pattern while swimming in the reservoir.}
\end{figure*}
\begin{figure*}[h]
 \centering
\includegraphics[width=16cm]{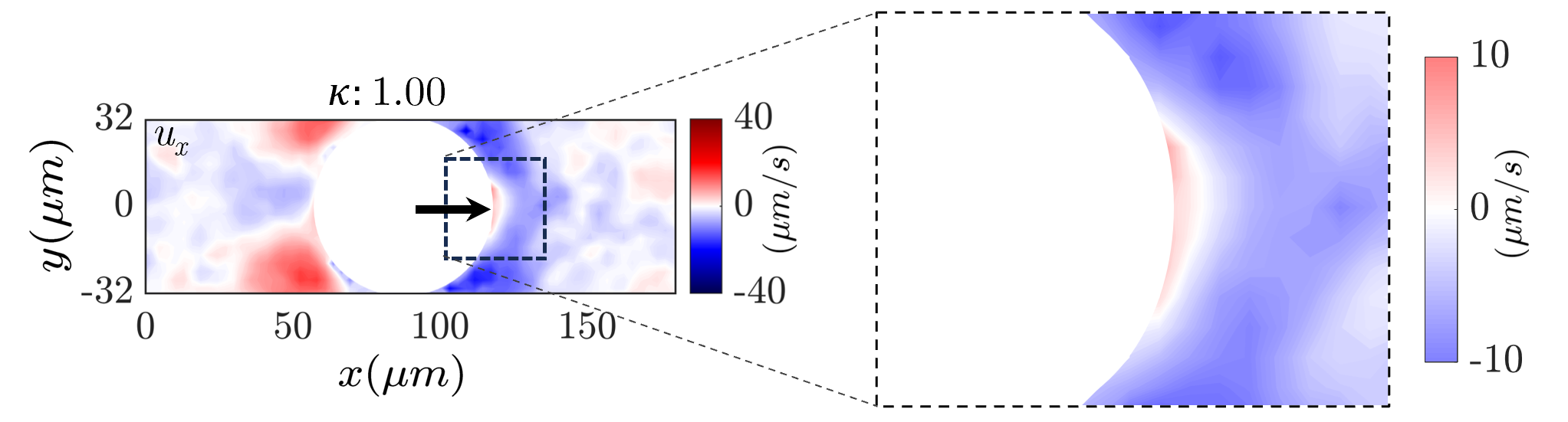}

 \caption{\label{fgr:12}Close-up view of the extreme front section of droplets for $u_x$ flow field for $\kappa=$ 1.00 (Regime-II). In this analysis, a smaller PIV mask was utilized to emphasize the slightly positive $u_x$ component at the extreme front. This finding indicates that the front portion of the droplet exhibits characteristics akin to a weak pusher under conditions of strong confinement.}
 \end{figure*}
 \begin{figure*}[h]
 \centering
 \includegraphics[width=16cm]{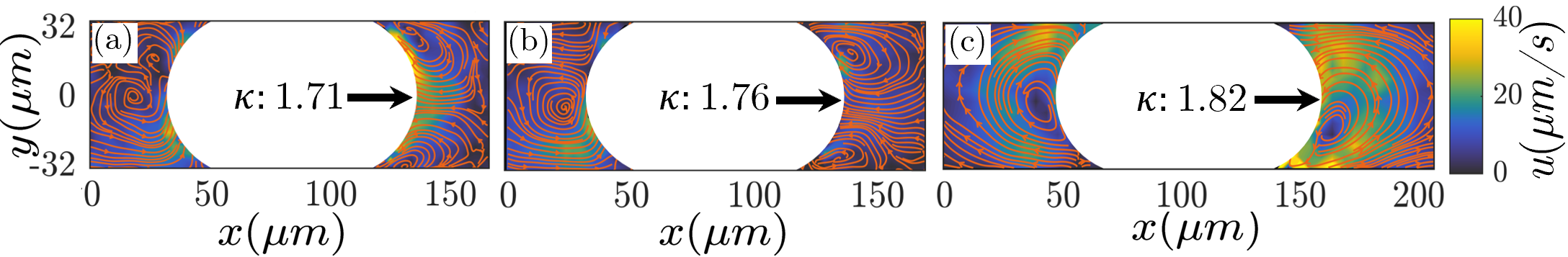}
 \caption{\label{fgr:13}The shifting of flow-field pattern from Regime-III to Regime-IV. \textbf{(a)} Droplet with $\kappa=1.71$ (Regime-III) and four circulation zones, \textbf{(b)} When $\kappa=1.76$ (Regime-III), one circulation from each front and back starts to push other and eventually tries to dominate, \textbf{(c)} With $\kappa=1.82$ (Regime-IV), the diminutive circulation zones vanish from both areas, resulting in a solitary circulation: one situated at the front and another at the rear.}
 \end{figure*}
 \begin{figure*}[h]
 \centering
 \includegraphics[width=16cm]{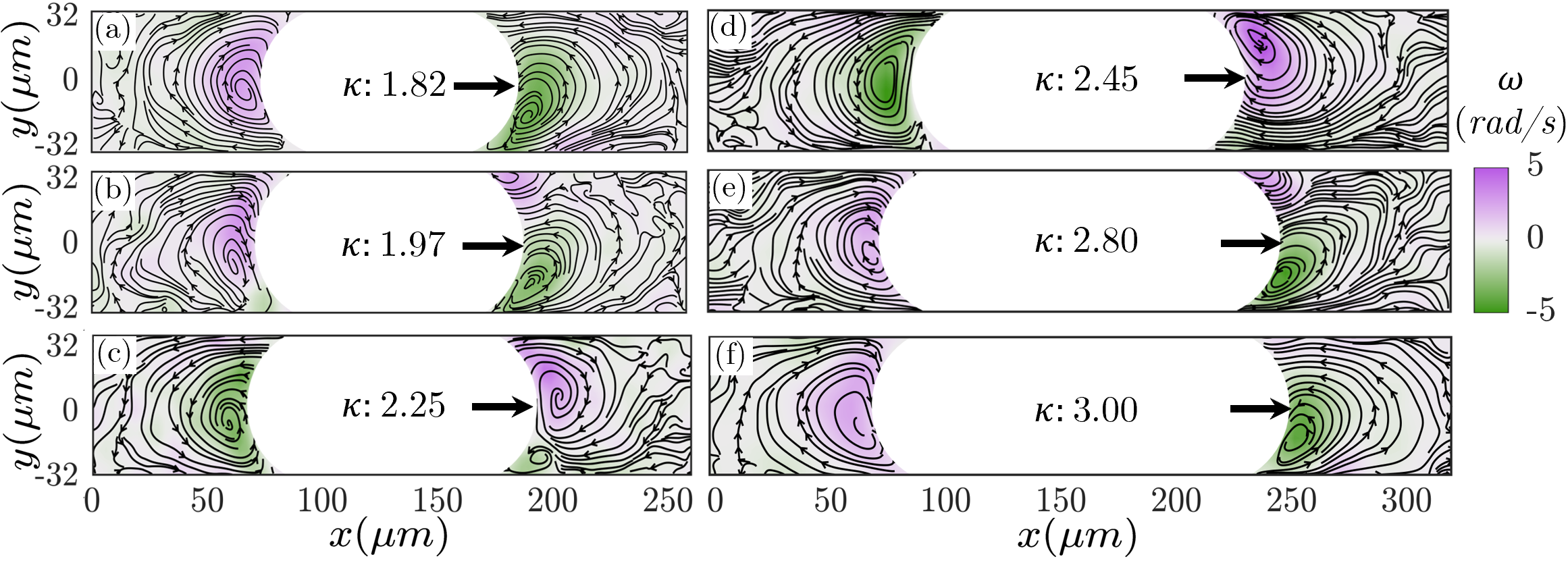}
 \caption{\label{fgr:14}In Regime-IV, a third small recirculation zone occasionally appears and disappears due to the strength of the orientation. In cases \textbf{(a)}, \textbf{(d)}, and \textbf{(f)} with $\kappa=1.82$, $2.45$, and $3.00$, respectively, that zone is absent. However, in cases \textbf{(b)}, \textbf{(c)}, and \textbf{(e)} with $\kappa=1.97$, $2.25$, and $2.80$, respectively, that zone reappears.}
\end{figure*}

 \begin{figure*}[h]
 \centering
\textbf{Videos of active droplets inside microchannel}
 \includegraphics[width=16cm]{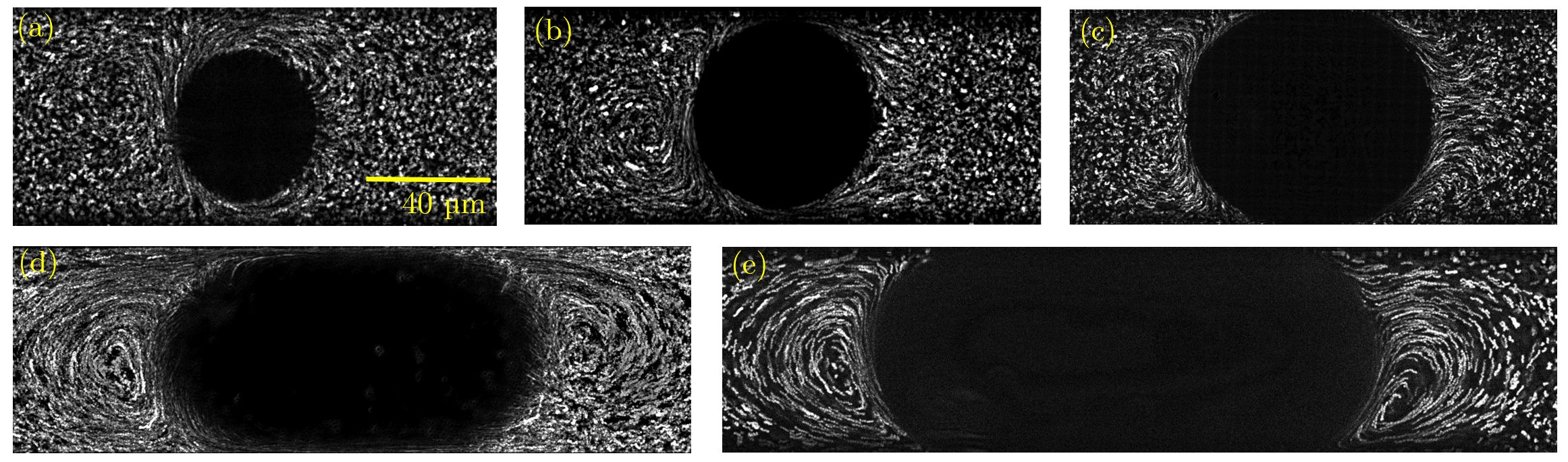}
  \centering \\  \raggedright(a) \textbf{Regime-I}: $\kappa$ $=0.85$ and $u_{d}$ $=20.12$ $\mu m/s$; for video click (\href{https://drive.google.com/file/d/1MZtsDDEAf_T1b6XbL9FY0-4lMKORynee/view?usp=drive_link}{\textbf{Video S1}}), \\(b) \textbf{Regime-II}: $\kappa$ $=1.00$ and $u_{d}$ $=17.05$ $\mu m/s$; for video click (\href{https://drive.google.com/file/d/106eoO-DSZNWl9EUMZXtIfpso6QQQlO6m/view?usp=drive_link}{\textbf{Video S2}}), \\(c) \textbf{Regime-III}: $\kappa$ $=1.50$ and $u_{d}$ $=14.78$ $\mu m/s$; for video click (\href{https://drive.google.com/file/d/1XRj5oaerT9ZIMk03pgk5mIWEljqDPLCG/view?usp=drive_link}{\textbf{Video S3}}). \\(d) \textbf{Regime-IV}: $\kappa$ $=2.25$ and $u_{d}$ $=10.34$ $\mu m/s$; for video click (\href{https://drive.google.com/file/d/1-BUTeEeCF0EB4oAooTPYzZymF14mBAhE/view?usp=drive_link}{\textbf{Video S4}}), \\(e) \textbf{Regime-IV}: $\kappa$ $=3.00$ and $u_{d}$ $=\;\;9.85$ $\mu m/s$; for video click (\href{https://drive.google.com/file/d/10WP7jf5wy1hJFn-4YadI1_aRvTjKB-5U/view?usp=drive_link}{\textbf{Video S5}}). \\The videos are processed and edited using ImageJ \cite{schneider2012nih} software with the FlowTrace extension \cite{Flow_Trace, Flow_Trace_2}.
\end{figure*}

\end{document}